\documentclass[journal]{IEEEtran}
\usepackage{graphicx}
\usepackage[cmex10]{amsmath}
\usepackage{subfigure}
\usepackage{amssymb}
\usepackage{algorithmicx}
\usepackage{algpseudocode}
\usepackage{algorithm}
\usepackage{color}
\usepackage{bigstrut}
\usepackage{booktabs}
\usepackage{balance}
\usepackage{stfloats}
\usepackage{placeins}
\usepackage{cite}
\usepackage{rotating}
\usepackage{enumitem}
\usepackage[colorlinks=false,linkcolor=black,
            allbordercolors={1 1 1},
            pdfborderstyle={/S/U/W 1},urlcolor={blue}]{hyperref}

\usepackage{multirow}
\usepackage{indentfirst}

\usepackage{acronym}
\usepackage[utf8]{inputenc}
\usepackage{makecell}

\acrodef{ecg}[ECG]{electrocardiogram}
\acrodef{emg}[EMG]{electromyogram}
\acrodef{pca}[PCA]{principal component analysis}
\acrodef{ste}[STE]{ST elevation}
\acrodef{std}[STD]{depression}
\acrodef{snrs}[SNRs]{signal-to-noise ratios}
\acrodef{snr}[SNR]{signal-to-noise ratio}
\acrodef{iqr}[IQR]{inter quartile range}
\acrodef{qtdb}[QTDB]{QT database}
\acrodef{prd}[PRD]{percentage root mean difference}
\acrodef{ldmfe}[LDMFE]{low-dimensional morphological feature extraction}
\acrodef{blw}[BLW]{baseline wander}

\newcommand{\IN}{\mathbb{N}}

\newcommand{\IR}{\mathbb{R}}

\newcommand{\bs}[1]{\boldsymbol{#1}}
\newcommand{\m}[1]{\mathbf{#1}}

\DeclareMathOperator{\EX}{\mathbb{E}}
\DeclareMathOperator*{\med}{median}
	
\ifCLASSINFOpdf
\else
\fi
\begin{document}
%
\title{ECG Beat Representation and Delineation by means of Variable Projection$^\dagger$}
%
%
%

\author{Carl~Böck$^*$,~
        Péter~Kovács$^*$,~
        Pablo Laguna,~
        Jens~Meier,~
        and~Mario~Huemer
\thanks{$^*$ These authors contributed equally to this work.}
\thanks{Carl Böck and Mario Huemer are with the Institute of Signal Processing, Johannes Kepler University Linz, Austria (email: \{carl.boeck,mario.huemer\}@jku.at).}
\thanks{Péter Kovács is with the Department of Numerical Analysis, Eötvös L. University, Budapest, Hungary (email: kovika@inf.elte.hu).}
\thanks{Carl Böck and Jens Meier are with the University Clinic of Anaesthesiology and Intensive Care, Kepler University Hospital, Johannes Kepler University Linz, Austria (email: \{carl.boeck,jens.meier\}@kepleruniklinkum.at).}
\thanks{Pablo Laguna is with the BSICoS group, University of Zaragoza, Zaragoza 50009, Spain, and also with the CIBER-BBN, Zaragoza 50018, Spain.}
\thanks{$^\dagger$ This work was supported by the Upper Austrian Medical Cognitive Computing Center (MC$^3$).}
\thanks{Manuscript received September 22, 2020; revised December 22, 2020, re-revised January 28, 2021, accepted January 31, 2021.}}

\maketitle

\begin{abstract}

\textit{Objective:} The electrocardiogram (ECG) follows a characteristic shape, which has led to the development of several mathematical models for extracting clinically important information. Our main objective is to resolve limitations of previous approaches, that means to simultaneously cope with various noise sources, perform exact beat segmentation, and to retain diagnostically important morphological information. \textit{Methods:} We therefore propose a model that is based on Hermite and sigmoid functions combined with piecewise polynomial interpolation for exact segmentation and low-dimensional representation of individual ECG beat segments. Hermite and sigmoidal functions enable reliable extraction of important ECG waveform information while the piecewise polynomial interpolation captures noisy signal features like the \ac{blw}.  For that we use variable projection, which allows the separation of linear and nonlinear morphological variations of the according ECG waveforms. The resulting ECG model simultaneously performs \ac{blw} cancellation, beat segmentation, and low-dimensional waveform representation. \textit{Results:} We demonstrate its  \ac{blw}  denoising and segmentation performance in two experiments, using synthetic and real data (Physionet QT database). Compared to state-of-the-art algorithms, the experiments showed less diagnostic distortion in case of denoising and a more robust delineation for the P and T wave. \textit{Conclusion:} This work suggests a novel concept for ECG beat representation, easily adaptable to other biomedical signals with similar shape characteristics, such as blood pressure and evoked potentials. \textit{Significance:} Our method is able to capture linear and nonlinear wave shape changes. Therefore, it provides a novel methodology to understand the origin of morphological variations caused, for instance, by respiration, medication, and abnormalities.

\end{abstract}

%
\IEEEpeerreviewmaketitle

\section{Introduction}
The \ac{ecg} is doubtlessly the most widely used biomedical signal for cardiac diagnosis. Usually, it is measured by recording the potential difference between electrodes placed on standardized locations on the surface of the body. The recorded traces consist of several deflections from the iso-electric level, which represent the individual waves that make up one heartbeat signal and consequently the \ac{ecg} (Fig.~\ref{fig:ecg_raw}). The time differences between individual waves, their duration, amplitude levels, polarity, and their shape all carry important clinical information \cite{sornmobook, ecg_book }. However, not only are these parameters relevant, their development over time, their beat-to-beat or long-term fluctuations, their responses to heart rate changes, and the interplay between them may also be of great clinical interest \cite{Laguna2016}. This gives raise to two major challenging tasks from a signal processing point of view: denoising and wave segmentation.

First and most obviously, redundant and noisy signal features should be eliminated while retaining clinically important information. For instance, the ECG is typically superimposed with \ac{blw}, which -- if not removed correctly -- interferes with correct diagnosis of specific illnesses. However, since \ac{blw} overlaps with the \ac{ecg} in the frequency domain, removing it may accidentally eliminate important diagnostic information \cite{sornmobook}. This would be critical, for instance in the case of ischemic ST-change detection, because corrupting the ST segment by removing the baseline could -- in the worst case -- lead to a wrong diagnosis \cite{Lenis2017}. Additionally, other noise sources, such as electromyographic noise and powerline interference, distort morphological features of the individual waves and should therefore be removed previous to further \ac{ecg} signal analysis. 

The next step in the analysis of \ac{ecg}s is typically the segmentation into individual beats and their three main deflections, that is P wave, QRS complex, and T wave (Fig.~\ref{fig:ecg_slices}). The wave segmentation is of high medical interest, since clinically relevant intervals, amplitude values, or other features of the individual waves and of the segments in between can be derived and used for diagnostics. Example applications are (ventricular) depolarisation assessment \cite{Romero2011}, quantification of QT variability \cite{Almeida2006} and ventricular repolarisation dispersion \cite{Arini2008, Arini2014, Laguna2016}. We have recently successfully used adaptive Hermite functions for \ac{ecg} wave segmentation, providing additionally a low-dimensional wave shape representation which potentially holds further important diagnostic information for the applications mentioned above \cite{Kovacs2017}. In general, Hermite functions have shown to be very well suited to \ac{ecg} signal processing, \ac{ecg} data compression \cite{hexp3, hexp5, Kovacs2020}, QRS complex clustering \cite{hexp4}, and detection of myocardial infarction \cite{hexp1}. Despite the usefulness of these functions some limitations remain, especially in relation to (baseline) noise and pathological ST segment elevations/depressions, which reveals the need to extend the basis function dictionary to achieve a correct beat delineation and representation. 
Addressing these limitations, we introduce two novel concepts for \ac{ecg} beat representation:

\begin{figure*}[!t]
\centering
\subfigure[Clinically relevant intervals and waves for \ac{ecg} strip with three cardiac cycles.]{
\includegraphics[width=5.5cm]{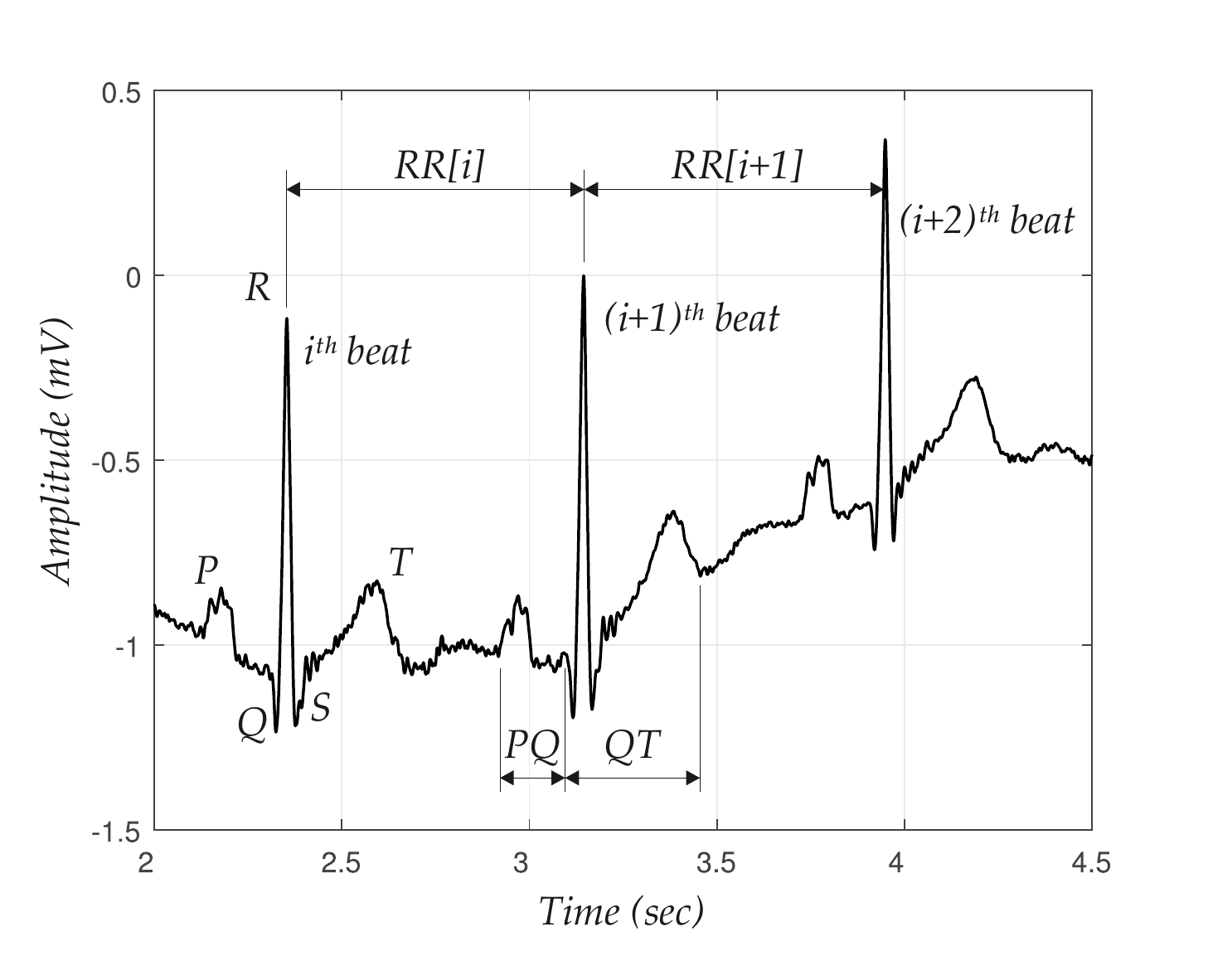}
\label{fig:ecg_raw}
}
\subfigure[Sliced ECG over time with segmented waves. Clinically useful parameters can be derived subsequently (e.g. PQ, QT)]{
\includegraphics[width=5.5cm]{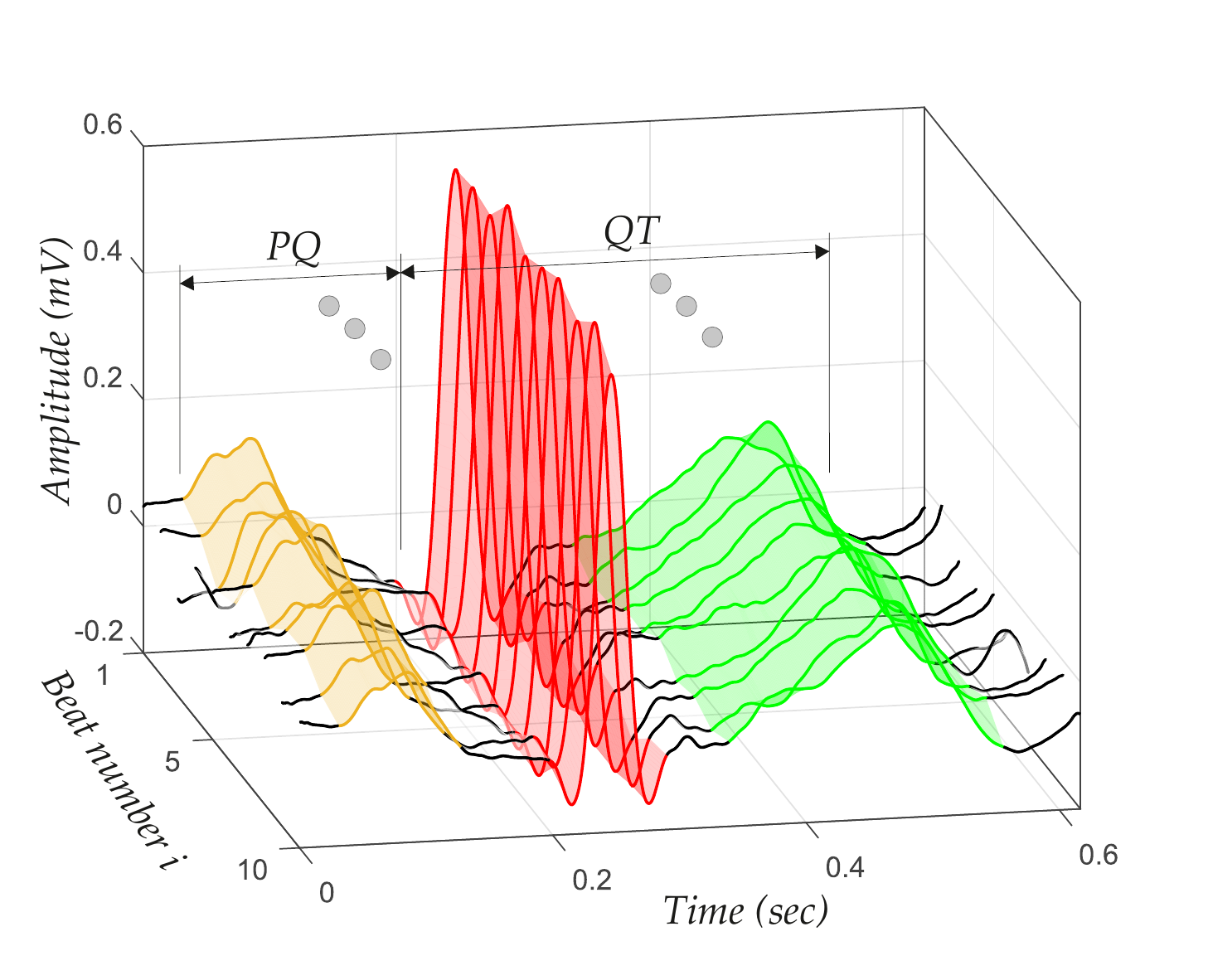}
\label{fig:ecg_slices}
}
\subfigure[Exemplary \ac{ecg} trace segmented into its single components as well as the unwanted baseline noise.]{
\includegraphics[width=5.5cm]{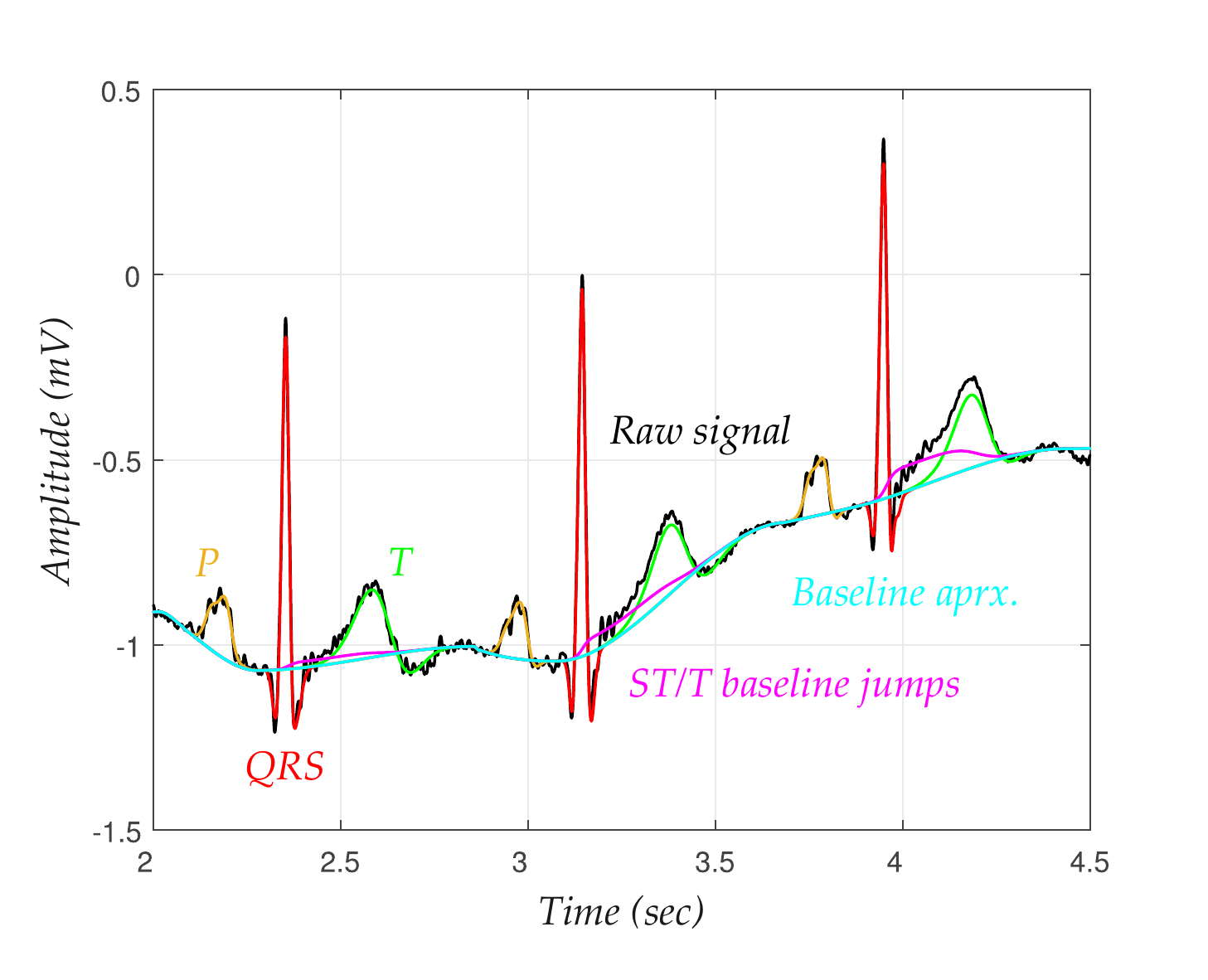}
\label{fig:ecg_aprx}
} 
\caption{Illustration of example \ac{ecg} traces with clinically relevant parameters and their segmentation into single waves.}
\label{fig_intro}
\end{figure*}

First, to improve baseline-noise-related limitations, we combine two well-known methodologies for \ac{ecg} signal analysis: spline interpolation for baseline estimation and adaptive Hermite functions for wave approximation and segmentation (Fig.~\ref{fig:ecg_aprx}). Note that state-of-the-art methods usually treat these two tasks separately, that means
 \ac{blw} removal is performed in a separate preprocessing step followed by morphological feature extraction \cite{Ansari2017, Lenis2017}. We, in contrast, simultaneously optimize the approximations of the single waves and of the baseline, which allows beats to be represented more accurately. In the case of \ac{blw} estimation, the PQ and TP segments are of particular interest, since they are assumed to be usually at the iso-electric level. These implicitly provided locations are used to identify the unwanted noisy baseline fluctuations by means of spline interpolation. A good baseline estimation, in turn, allows better beat approximation, which generally leads to improved \ac{ecg} signal representation.

Our second novel concept extends the basis function dictionary to include sigmoidal atoms, that model possible (pathological) intra-beat baseline jumps, such as an \ac{ste} or \ac{std}. Based on these two major conceptual novelties, noisy and redundant signal features are minimized while important morphological and diagnostic information is extracted for the segmented wave components.

We evaluated and compared our method to state-of-the-art techniques in two independent experiments. First, to show that our method removes \ac{blw} efficiently without distorting important morphological features, we compared our results to those of Lenis et al., who analyzed the most prominent \ac{blw} removal techniques in a recent simulation study \cite{Lenis2017}. Second, we used the well-known Physionet \ac{qtdb} \cite{PhysioNet, Laguna1997} to evaluate the effectiveness of our method in \ac{ecg} wave segmentation, as it provides a good benchmark for this task. 

This paper comprises six sections. Sec.~\ref{sec:waveform_modeling} describes the nonlinear waveform modeling of \ac{ecg} beats, while Sec.~\ref{sec:dictionary} elaborates on the construction of the dictionary. Section~\ref{sec:constraints} defines the constraints used for optimization. Our method's ability to perform \ac{blw} removal and \ac{ecg} wave segmentation is illustrated in Sections~\ref{sec:baseline} and \ref{sec:wave_segmentation}. Section~\ref{sec:conclusion} concludes our work, emphasizing the strengths of our method and providing possible future applications.

\section{Nonlinear waveform modeling}
\label{sec:waveform_modeling}
Due to their simplicity and low computational complexity, linear models are frequently used to model \ac{ecg} signals \cite{addison2005, castells2007, Kovacs2020}. One of the main challenges in this context is to find a proper set of atoms (i.e., a dictionary) that matches the problem area. Since the main \ac{ecg} waveshapes are influenced by many person-specific factors such as age, gender, morbidities, and medications, no dictionary exists that is optimal for all cases. Hence, we developed a nonlinear least-squares model, that is tailored precisely to a single person. In order to track morphological evolvement over time, the person-specific model is then readjusted beat by beat by means of local nonlinear optimization. 

Let us consider an analog signal $f(t)$ which is sampled at time instances $t_1,\ldots,t_N$. The nonlinear approximation of the observed data is then given as
\begin{equation}
	f(t_n)\approx\eta(\m{c},\bs\alpha; t_n)=\sum_{j=0}^{J-1} c_j \varphi_j(\bs\alpha;t_n)\quad (n=1,\ldots,N)\,,
\label{eq:nonlinmodel}
\end{equation}
where $\left\{\varphi_j(\bs\alpha;\cdot)\,|\,0\leq j < J\right\}$ denotes the dictionary, $\bs\alpha\in\IR^m$ is the vector of parameters controlling the nonlinearities in the signal modeling, and $\m{c} = \left[c_0, c_1, \ldots c_{J-1}\right]^T$ is the vector of corresponding coefficients, determined by solving a simple linear least-squares problem for a given $\bs\alpha$. This 
can be written in matrix-vector form $\m{f}\approx\bs{\Phi}\left(\bs\alpha\right)\m{c}$, where $\m{f}=\left[f(t_1),\ldots,f(t_N)\right]^T$, and $\left\{\bs\Phi\left(\bs\alpha\right)\right\}_{n,j}=\varphi_j(\bs\alpha;t_n)$ for $n=1,\ldots,N,\;j=0,\ldots,J-1$. 

In this work, a heartbeat representation comprises four components: the QRS, T and P waves and the baseline. Each of these components is modeled by an individual nonlinear model, and their sum defines the joint model of the heartbeat: 
\begin{equation}
	\m{f}\approx \bs{\eta}^{\text{QRS}} + \bs{\eta}^{\text{T}} + \bs{\eta}^{\text{P}} + \bs{\eta}^{\text{BL}}\,.
\label{eq:joinmodel}
\end{equation}
Note that each component has its own dictionary and linear ($\m{c}$) and nonlinear ($\bs\alpha$) parameters (see Table~\ref{tab:dictpars}).

In order to find the best parameters for a given heartbeat signal, we consider the following optimization problem:
\begin{equation}
	\min_{\bs{\alpha}} r_2(\bs{\alpha})=\min_{\bs{\alpha}} \left\|\m{f} - \bs{\Phi}(\bs{\alpha})\bs{\Phi}^+(\bs{\alpha})\m{f}\right\|_2^2\,,
\label{eq:varpro}
\end{equation}
where $\bs{\Phi}^+(\bs{\alpha})$, which is equal to $\bs{\Phi}^T(\bs{\alpha})$ if $\varphi_j(\bs\alpha;t_n)$ form an orthonormal system, denotes the Moore--Penrose pseudoinverse of the matrix $\bs{\Phi}(\bs{\alpha})$, and
\begin{align}
	\bs{\alpha}&=\left(\bs{\alpha}^{\text{QRS}}, \bs{\alpha}^{\text{T}}, \bs{\alpha}^{\text{P}}, \bs{\alpha}^{\text{BL}}\right)^T\,,\\
	\bs{\Phi}(\bs{\alpha})&=
	\left(
		\bs{\Phi}^{\text{QRS}}(\bs{\alpha}^{\text{QRS}}),\bs{\Phi}^{\text{T}}(\bs{\alpha}^{\text{T}}),\bs{\Phi}^{\text{P}}(\bs{\alpha}^{\text{P}}),\bs{\Phi}^{\text{BL}}(\bs{\alpha}^{\text{BL}})
	\right)
\end{align}
are the column-wise concatenations of the components' nonlinear parameters and the corresponding dictionaries. Note that, for a given $\bs{\alpha}$, the linear parameters are calculated according to the least-squares solution $\m{c}(\bs{\alpha}):=\bs{\Phi}^+(\bs{\alpha})\m{f}$. Therefore, only the vector of nonlinear parameters $\bs{\alpha}$ is to be optimized. The so-called variable projection (VP) functional $r_2$ was introduced by Golub and Pereyra \cite{golub_pereyra1973}. As they provided an exact formula for the gradient, local search techniques, such as trust-region methods \cite{varpro_matlab}, can be applied to minimize $r_2$. This approach has a wide range of applications  \cite{golub_pereyra2003}. Based on this general nonlinear waveform model, we describe the atoms we selected to represent the \ac{ecg} signal.

\begin{table}[!t]
\caption{Parameters of the dictionaries. Note that $\tau_{\min},\tau_{\max}$ denote the min./max. distance to the initially detected R peak $R_\text{init}$.}
\label{tab:dictpars}
\centering
\scalebox{0.73} { { \renewcommand{\arraystretch}{1.3} 
  \begin{tabular}{| c || c | c | c | c |}
		\hline
		& $\bs{\eta}^{\text{QRS}}$ & $\bs{\eta}^{\text{T}}$ & $\bs{\eta}^{\text{P}}$ & $\bs{\eta}^{\text{BL}}$\\ \hline\hline
		atom & \begin{tabular}{@{}c} Hermite \\ sigmoid \end{tabular} & \begin{tabular}{@{}c} Hermite \\ sigmoid \end{tabular} & Hermite & \begin{tabular}{@{}c} Piecewise \\ polynomials \end{tabular}\\ \hline
		$J$ & 7+1 &  4+1 & 4 &1 \\ \hline
		$\m{c}$ & least squares & least squares & least squares & interpolation \\ \hline
		$\bs{\alpha}$ & $\lambda_{\text{QRS}},\tau_{\text{QRS}}$ & $\lambda_{\text{T}},\tau_{\text{T}}$ & $\lambda_{\text{P}},\tau_{\text{P}}$ & \begin{tabular}{@{}c} $\lambda_{\text{QRS}},\tau_{\text{QRS}},$ \\ $\lambda_{\text{T}},\tau_{\text{T}}$ \end{tabular} \\ \hline
		$[\lambda_{\min},\lambda_{\max}]$ (1/s) & $[44.12,85.71]$ & $[14.78,30.61]$ & $[39.47,68.18]$ & -- \\ \hline
		$[\tau_{\min},\tau_{\max}]$ (ms) & $[-68,68]$ & $[133,343]$ & $[-R_\text{init}+44,-112]$ & -- \\ \hline
  \end{tabular} 
}}
\end{table}

\section{Constructing the dictionary}
\label{sec:dictionary}

A good dictionary matches the main characteristics of the modelled signal; this means that the dictionary atoms are highly correlated with the main components of the observed data. Below we describe how we developed a dictionary that is tailored specifically to the problem of \ac{ecg} beat segmentation / representation (Sec.~\ref{sec:dict_beat}) and simultaneously capturing the noisy baseline information (Sec.~\ref{sec:dict_bl}).

\subsection{Dictionary for the QRS, T and P waves} \label{sec:dict_beat}
Nonlinear parametric modeling of \ac{ecg} waveforms dates back to the 1980s, when S\"ornmo et al.\ \cite{hexp2} introduced the dilated Hermite functions for approximating the shape of the QRS complex. Their work inspired many others: Laguna et al.\ \cite{hexp3} and Kov\'acs et al. \cite{Kovacs2020} compressed \ac{ecg} signals; Lagerholm et al.\ \cite{hexp4} utilized this approach for heartbeat clustering purposes; Haraldsson et al.\ \cite{hexp1} extracted features by means of Hermite expansion to detect myocardial infarction, while Sandryhaila et al.\ \cite{hexp5} improved reconstruction accuracy by using the discrete analogue of the dilated Hermite function system. These approaches utilize a single nonlinear parameter that is the dilation of atoms. In \cite{hexp6}, a translation parameter was added, which we used in previous work to develop an \ac{ecg} segmentation algorithm \cite{Kovacs2017}.

The shape similarity between Hermite functions and \ac{ecg} waveforms (P-QRS-T) explains the suitability of the former to represent the latter. We derive the family of Hermite functions as
\begin{equation}
	\varphi_j(t) = h_j(t)\cdot \exp({-t^2/2}) \cdot 2^{j/2} \big/ \sqrt{\pi^{1/2} j!}\textrm{,} \qquad (j\in\IN)\textrm{,}
	\label{eq:hermitedef}
\end{equation}
where $h_j$ denotes the well-known Hermite polynomials with $h_0(t)=1,\, h_1(t)=2t$ and
\begin{equation*}
	h_{j+1}(t) = 2 t h_j(t)-2 j h_{j-1}(t)\textrm{,} \qquad (j\in\IN^+)\textrm{.}
\end{equation*}
It is well known that this family of functions $\left\{\varphi_j\,:\,j\in\IN\right\}$ forms an orthonormal and complete system in $L^2(\IR)$ with respect to the usual scalar product and norm \cite{szego}.

\begin{figure}[!t]
\centering
  \subfigure[Unscaled Hermite functions.]{
  \includegraphics[scale=0.332, trim=155 250 120 250, clip]{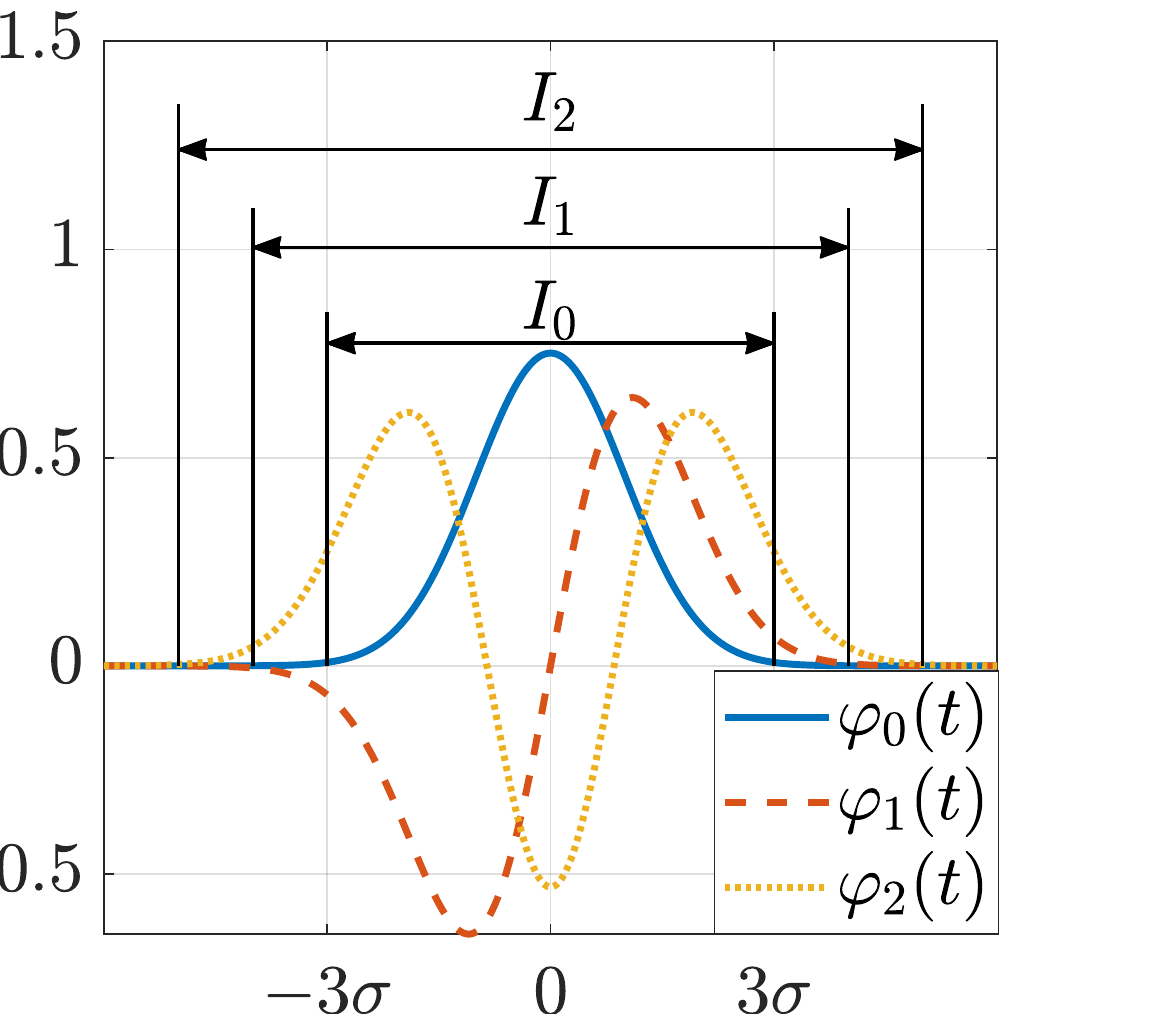}
  \label{fig:unscaled_hermite}
  } 
  \subfigure[Scaled Hermite functions.]{
  \includegraphics[scale=0.322, trim=155 250 120 250, clip]{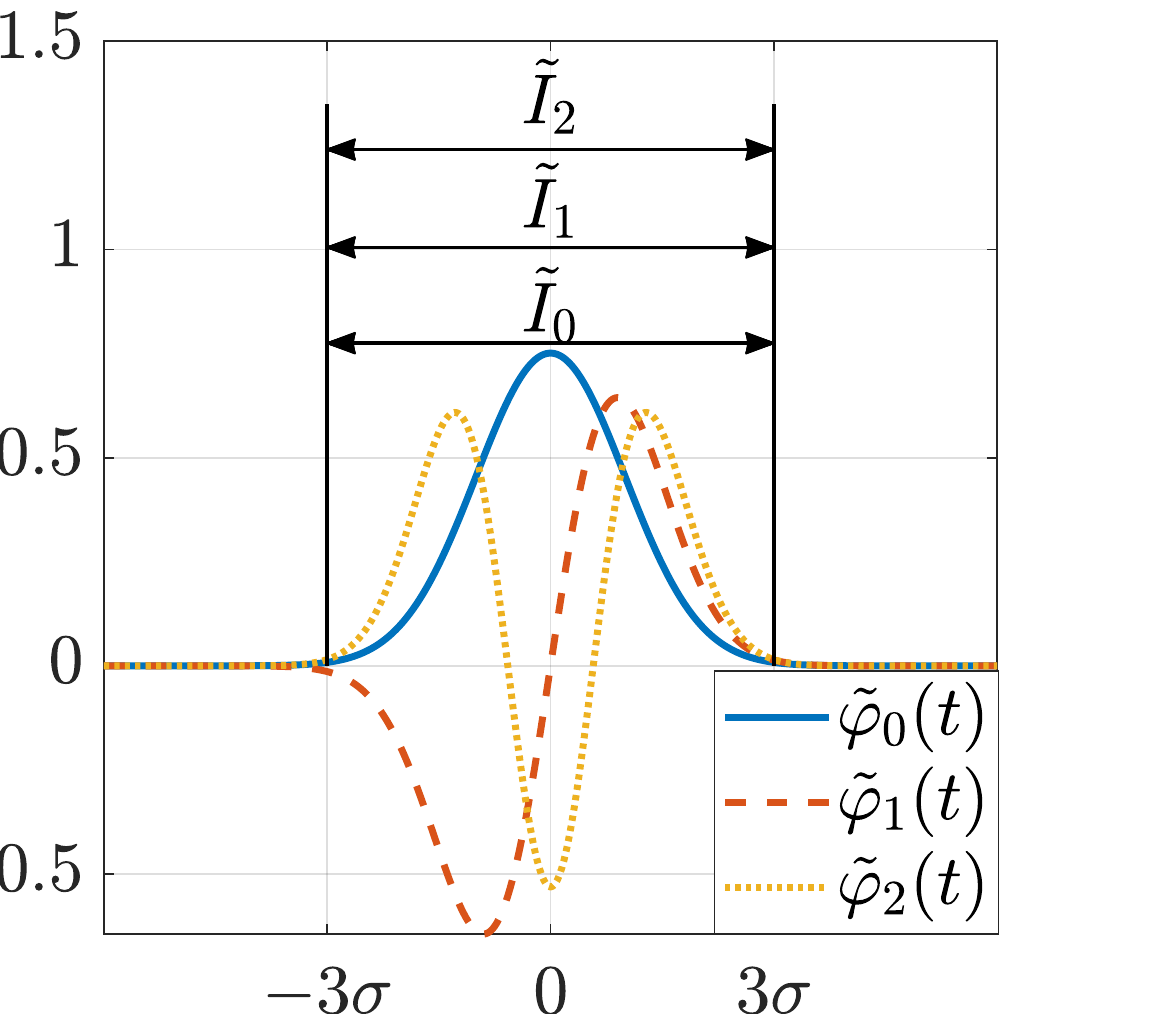}
  \label{fig:scaled_hermite}
  } 
\caption{Scaling of the Hermite functions.}
\label{fig:scaling}
\end{figure}

A very useful property of the Hermite functions is that they are localized in time, which means that $\lim_{t\rightarrow \pm \infty} \varphi_j(t)=0$ for all $j\in\IN$. It can be shown that a series of nested intervals $I_j\subset I_{j+1}$ exists in which the corresponding $\varphi_j$ functions are significantly different from zero; outside of these intervals these functions decrease rapidly due to the exponential factor in \eqref{eq:hermitedef}. This is illustrated in Fig.~\ref{fig:unscaled_hermite}, where we show the first three elements of the family of Hermite functions. Clearly, the individual wave approximations $\bs{\eta}^{\text{QRS}}, \bs{\eta}^{\text{T}}, \bs{\eta}^{\text{P}}$ have no effect outside the corresponding $I_j$ intervals. In order to improve on the separation of the QRS, T and P components from our previous approach \cite{Kovacs2017}, we unify the $I_j$ intervals by rescaling the Hermite functions as follows: 
\begin{equation}
	\tilde{\varphi}_j(t)=\varphi_j(1.11^j \cdot t) \qquad (j\in\IN)\,;
\label{eq:rescaledfuns}
\end{equation}
we found the value of the scaling factor ($1.11$) experimentally such that the resulting $\tilde{I}_j$ intervals are approximately equal to $\tilde{I}_0$ (Fig.~\ref{fig:scaled_hermite}). Note that  $\varphi_j(\bs\alpha;t_n)$ do not form an orthogonal system any more, however allow better separability of the QRS, T and P components. In Sec.~\ref{sec:constraints}, we additionally provide medical constraints on the length of $\tilde{I}_0$, which now apply to the whole family of the rescaled Hermite functions.

\begin{figure}[!t]
\centering
  \subfigure[Beat appr. without sigmoid.]{
  \includegraphics[scale=0.3]{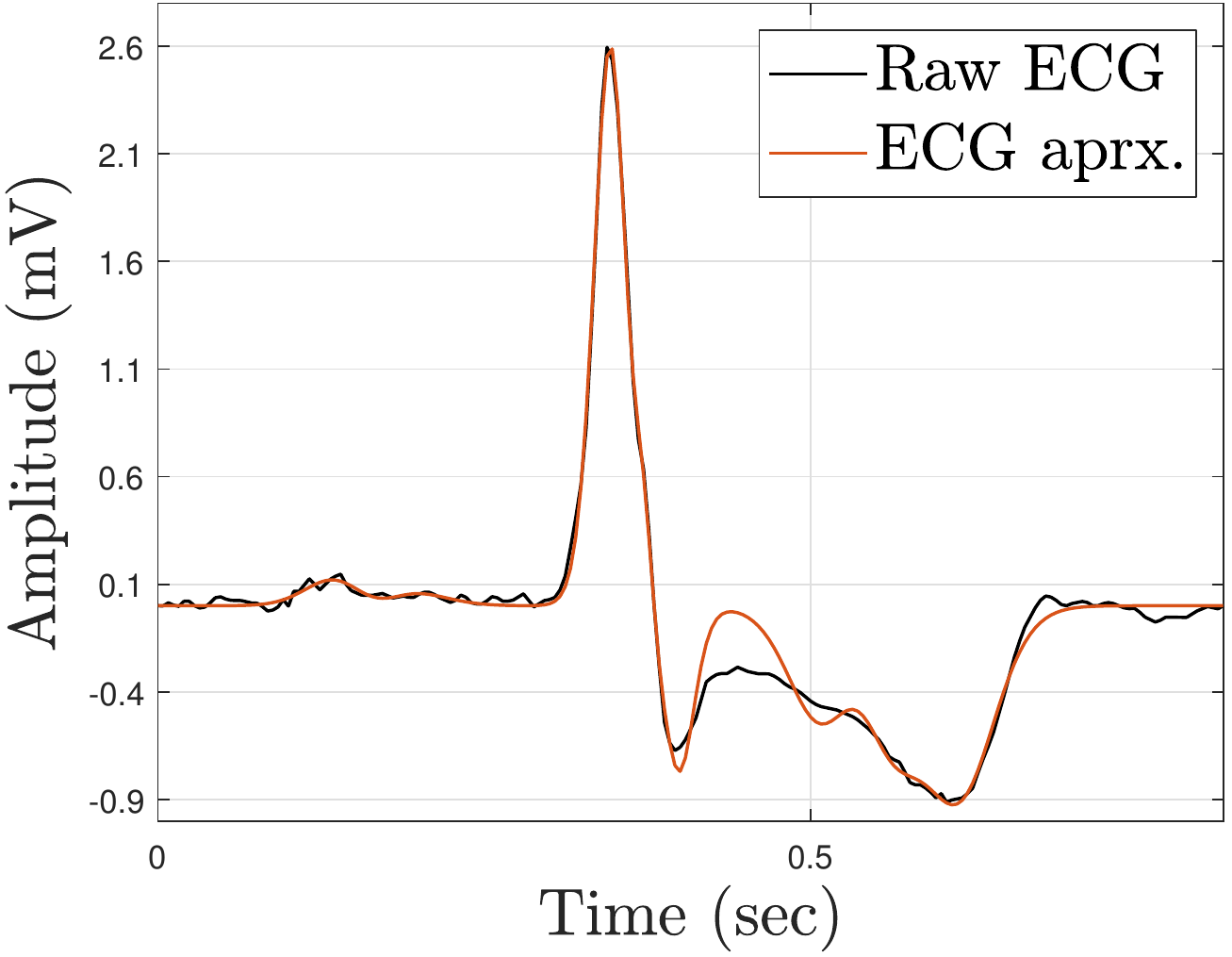}
  \label{fig:beat_aprx_no_sig}
  } 
  \subfigure[Wave segm. without sigmoid.]{
  \includegraphics[scale=0.3]{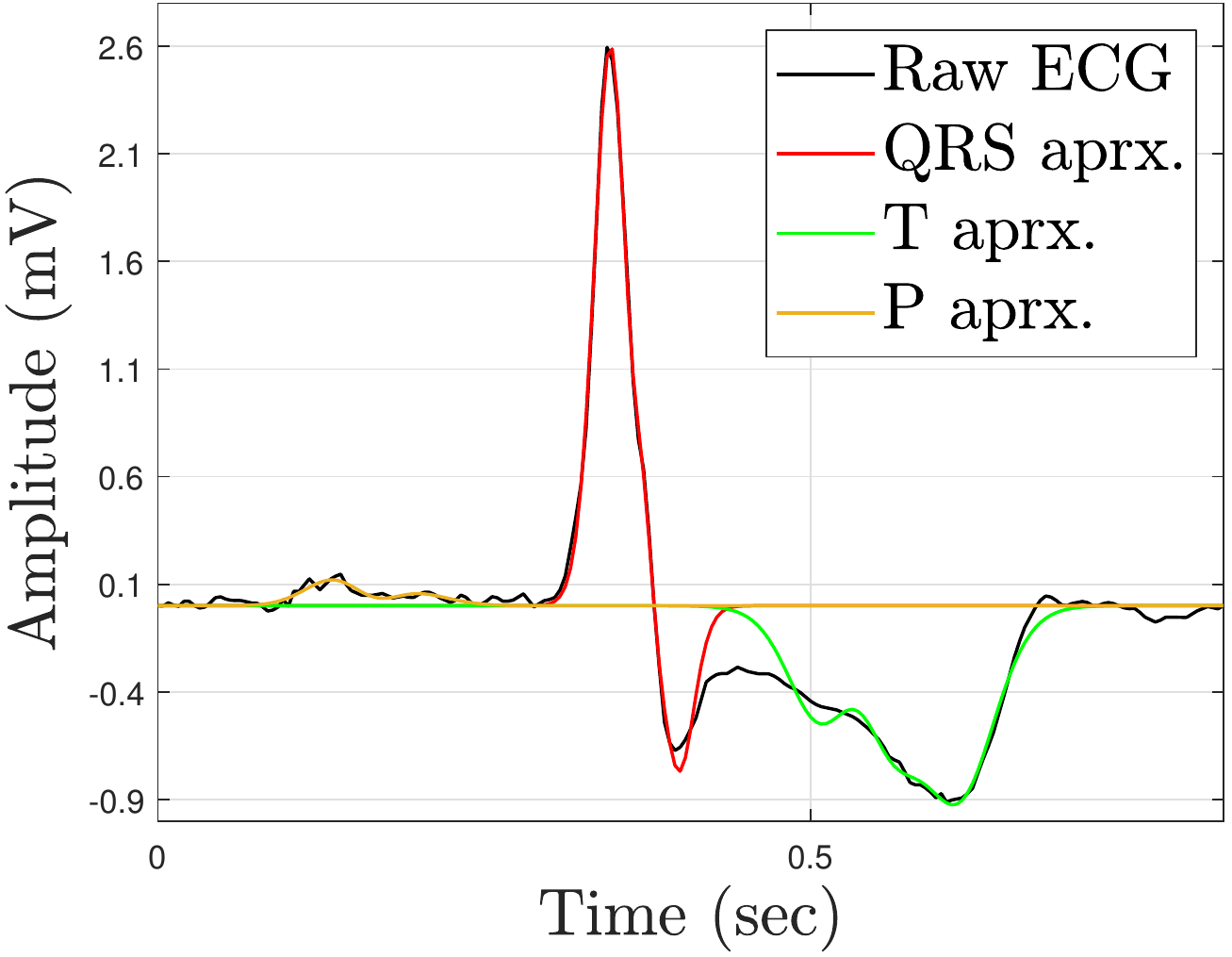}
  \label{fig:wave_aprx_no_sig}
  } 
  \subfigure[Beat appr. with sigmoid.]{
  \includegraphics[scale=0.3]{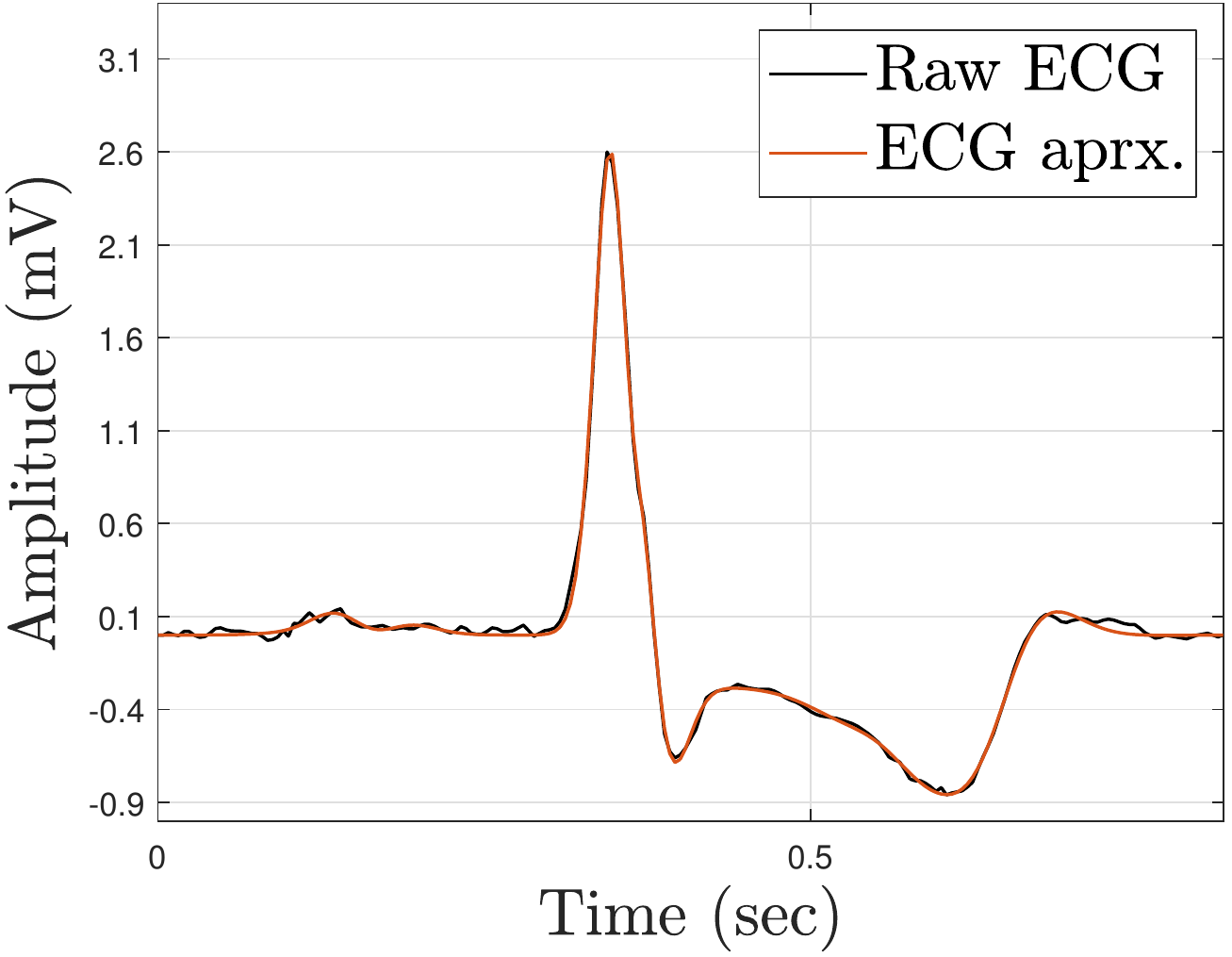}
  \label{fig:beat_aprx_w_sig}
  } 
  \subfigure[Wave segm. with sigmoid.]{
  \includegraphics[scale=0.3]{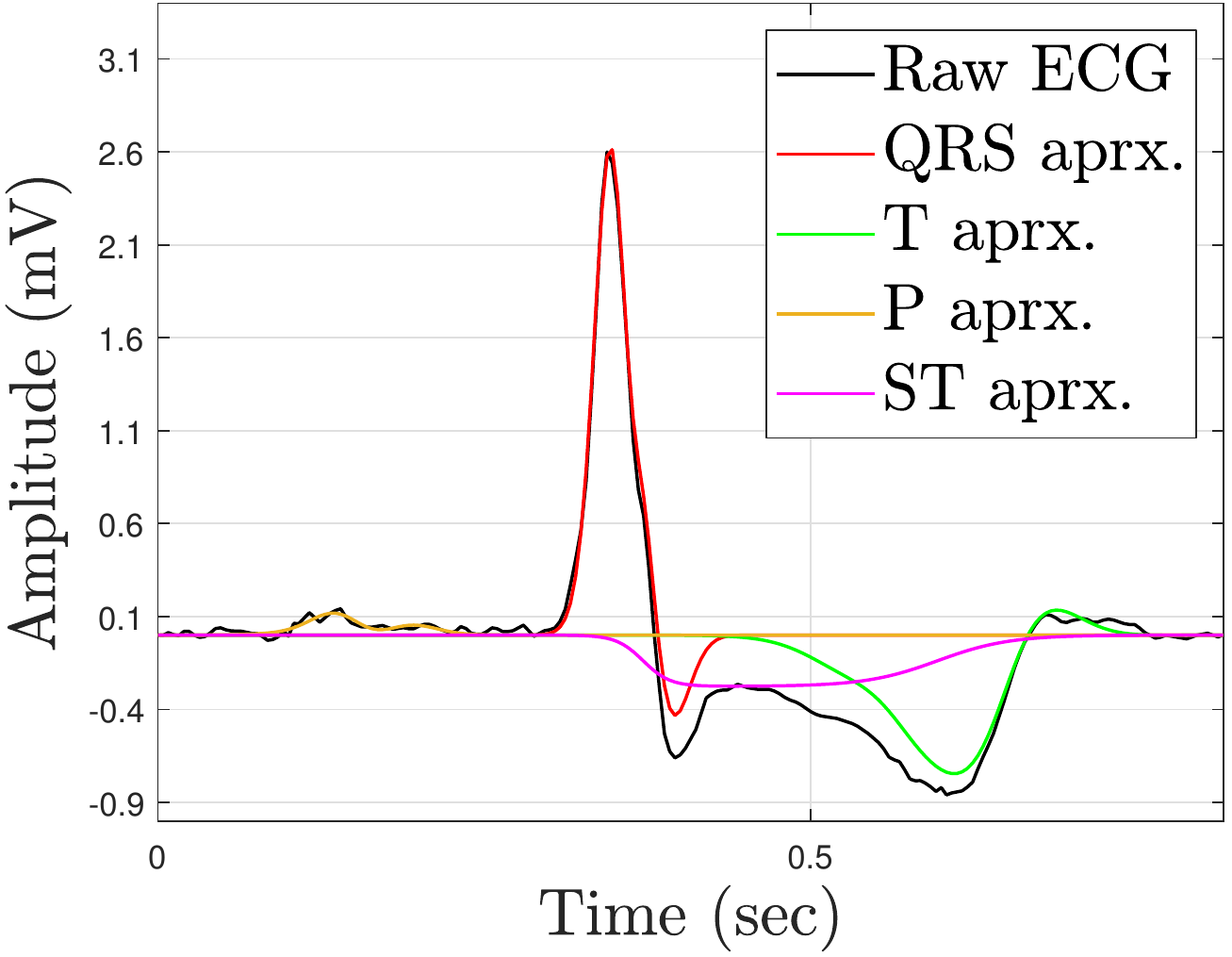}
  \label{fig:wave_aprx_w_sig}
  }   
\caption{Demonstrating the advantages of a sigmoid atom.}
\label{fig:scaling}
\end{figure}

Although, the system of Hermite functions serves as a foundation of our dictionary, we found that it is not sufficient to represent the most common waveshapes in \ac{ecg}s. In fact, in a preliminary experiment conducted together with medical experts, we observed that specific physiological and pathological baseline jumps (e.g., \ac{ste}/\ac{std}) are not captured well by Hermite functions. We illustrate this phenomenon in Fig.~\ref{fig:beat_aprx_no_sig}-\ref{fig:wave_aprx_no_sig}, which shows a significant difference between the signal levels of each side of the QRS complex. However, this component can be represented very well by a sigmoid function:
\begin{equation*}
	s(t) = 1/(1+\exp(-2t)) \qquad (t\in\IR)\textrm{,}
\end{equation*}
as shown in  Fig.~\ref{fig:beat_aprx_w_sig}-\ref{fig:wave_aprx_w_sig}. The sigmoid functions are aligned with the Hermite atoms and compensate each other across a beat. Consequently, we extend the system of Hermite functions to include the logistic sigmoid curve $s(t)$. This allows us to represent a wider class of clinically relevant \ac{ecg} waveforms. 

Relative position ($\tau$) and the width ($\lambda$) of the QRS, T and P waves change dynamically from beat to beat for physiological reasons such as respiration. It therefore is reasonable to parametrize the dictionary in order to adjust its atoms to the current heartbeat. Building upon former work \cite{hexp6, Kovacs2017}, we use an affine argument transform to parametrize the rescaled Hermite and the sigmoid functions: 
\begin{align*}
	\tilde{\varphi}_j(\lambda,\tau; t) &= \tilde{\varphi}_j( \lambda\cdot(t-\tau) )\textrm{,}\\
	s(\lambda,\tau; t)&=s( \lambda\cdot(t-\tau) )\textrm{,} \qquad (t,\tau\in\IR,\,\lambda\in\IR^+)\textrm{.}
\end{align*} 
These functions can be uniformly sampled at time $t_1,\ldots,t_N$, which allows the corresponding adaptive dictionary to be defined as follows: 
\begin{equation*}
	\bs{\Phi}^{\text{QRS}}(\bs{\alpha}^{\text{QRS}}) = \big( \bs{\tilde{\varphi}}_0(\bs{\alpha}^{\text{QRS}}),\ldots,\bs{\tilde{\varphi}}_{J-1}(\bs{\alpha}^{\text{QRS}}),\m{s}(\bs{\alpha}^{\text{QRS}})\big)\textrm{,}
\end{equation*}  
where $\bs{\bs{\alpha}^{\text{QRS}}}=(\lambda_{\text{QRS}},\tau_{\text{QRS}})$. Hence, we use the first seven rescaled Hermite atoms and the sigmoid atom for modeling the QRS complex. The dictionaries for the T and the P waves are defined analogously, using the first four rescaled Hermite atoms and additionally the sigmoid atom in case of the T wave. Table~\ref{tab:dictpars} summarizes the parameter setup of each component. Note that we do not use the sigmoid function in $\bs{\Phi}^{\text{P}}$, assuming that there is no physiological / pathological baseline jump between onset and end of the P wave. This decision was made based on preliminary experiments and in agreement with medical experts. Hence, we assume that differences in the amplitude levels on either side of the P wave originate from unwanted \ac{blw}, which is modelled as described in the following section. 

\subsection{Dictionary for the baseline wander} \label{sec:dict_bl}

\ac{blw} is a low-frequency noise caused mainly by respiration and body movements. In order to cancel out unwanted \ac{blw}, polynomial fitting and subtraction \cite{sornmobook, ecg_book} have proved to be very efficient approaches, where the baseline is estimated by piecewise polynomial interpolation. However, the performance of these techniques heavily depends on the nodes, which are typically located at the estimated time instants of the PQ and/or TP segments, since these are assumed to be at the iso-electric level. 

\begin{figure}[!t]
\centering
  \subfigure[Baseline approximation for a single beat.]{
  \includegraphics[scale=0.3]{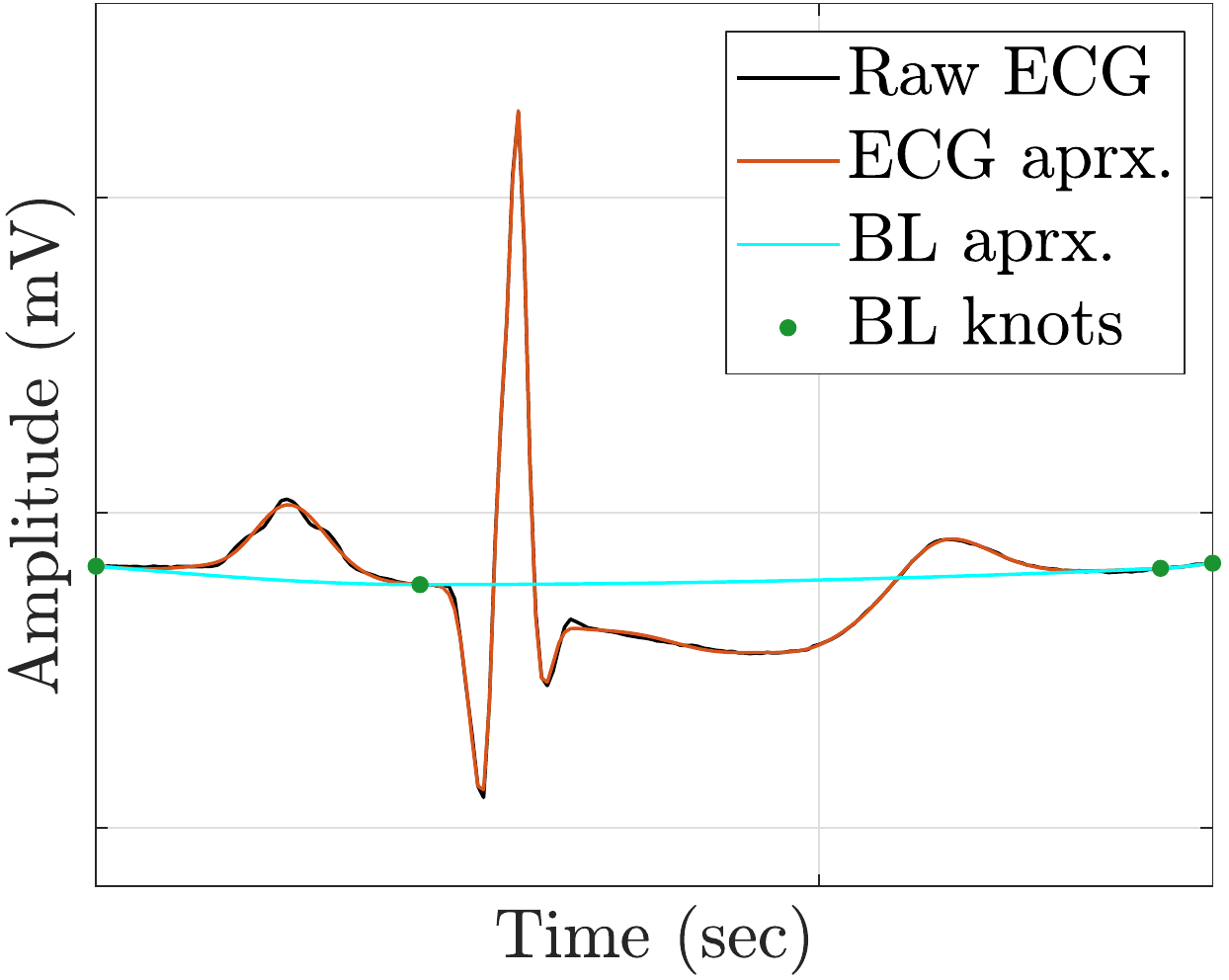}
  \label{fig:bl_single_beat}
  } 
  \subfigure[Baseline approximation for the ECG recording.]{
  \includegraphics[scale=0.3]{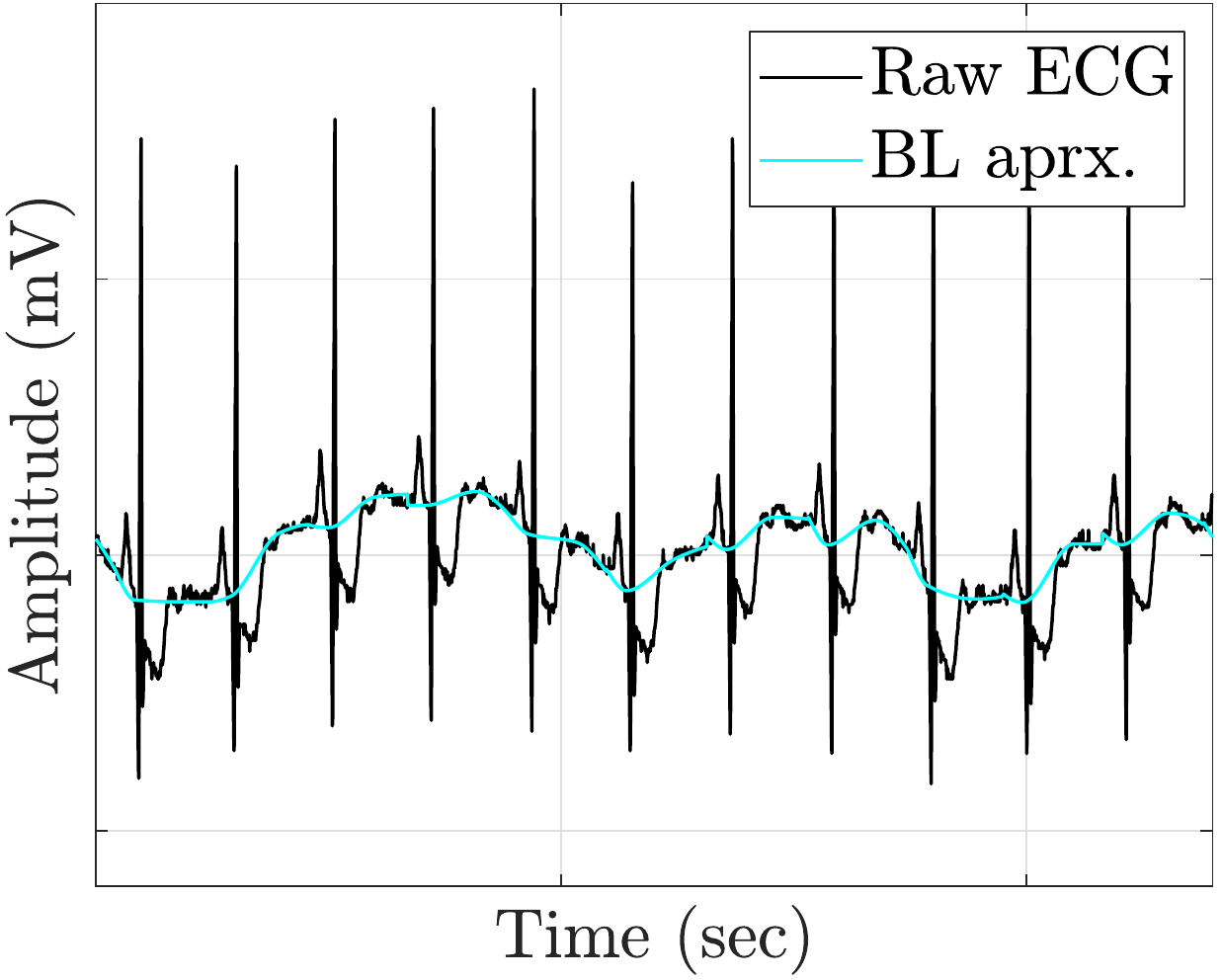}
  \label{fig:bl_more_beats}
  } 
\caption{Baseline estimation using the proposed method.}
\label{fig:bl}
\end{figure}

We improve on this baseline estimation approach by automatically setting the nodes via the VP optimization stated in \eqref{eq:varpro}. More specifically, as illustrated in Fig.~\ref{fig:bl_single_beat}, we define $4$ knots for each heartbeat as follows:
\begin{itemize}
	\setlength\itemsep{0.5em}
	\item{\makebox[2.4cm]{boundary points:\hfill} $x_1=1, \; x_4=N;$}
	\item{\makebox[2.4cm]{fiducial points:\hfill} $x_2=\tau_{\text{QRS}}-4/\lambda_{\text{QRS}}, \; x_3=\tau_{\text{T}}+4/\lambda_{\text{T}},$}
\end{itemize}
where $N$ denotes the number of samples representing one beat. The second and the third knot depend on the affine parameters of the QRS complex and the T wave, respectively. To derive these formulas, we apply the well-known three-sigma rule to the first Hermite function (see, e.g., Section~\ref{sec:constraints}). Here, $\tau_{\text{QRS}}-3/\lambda_{\text{QRS}}$ can be interpreted as the starting point of the QRS approximation $\bs{\eta}^{\text{QRS}}$, and $x_2$ is a good estimate of the PQ segment location. Since the same is true for the T wave, where $\tau_{\text{T}}+3/\lambda_{\text{T}}$ corresponds to the end of the T wave, we assume that $x_4$ is, again, at the iso-electric level. In each iteration of the optimization we first determine the knots and then compute the corresponding piecewise cubic Hermite interpolation polynomial (pchip). In order to prevent superfluous oscillation of the baseline approximation, we use shape-preserving polynomial interpolation \cite{monspline}. For a given parameter vector $\bs{\alpha}=(\lambda_{\text{QRS}},\tau_{\text{QRS}},\lambda_{\text{T}},\tau_{\text{T}})$, the fitted polynomial curve $p(\bs{\alpha};t)$ is the only atom in the baseline dictionary:
\begin{equation*}
	\left\{\bs{\Phi}^{\text{BL}}(\bs\alpha)\right\}_i = \left\{ \m{p}(\bs\alpha)\right\}_i=p(\bs{\alpha};t_i) \qquad (i=1,\ldots,N)\textrm{.}
\end{equation*}
An example of the resulting baseline interpolation is illustrated in Fig.~\ref{fig:bl_more_beats}, which qualitatively shows that the \ac{blw} is captured very well. A quantitative analysis follows in Section~\ref{sec:baseline}. 

\section{Constraining the optimization}
\label{sec:constraints}

Once the optimal parameters have been found, the basic segmentation of the heartbeat is given by the components of the nonlinear model in \eqref{eq:joinmodel}. Note that we optimize the nonlinear parameter vector $\bs{\alpha}$ in \eqref{eq:varpro} by minimizing the least-squares error of the sum of each components' approximation. Therefore, it may happen that the error of the approximation is small, although the delineation itself is inaccurate. Our main objective here is to provide constraints which force the optimization to find a good approximation with compactly supported nonlinear components. Mathematically speaking, the almost-orthogonal approximations $\bs{\eta}^{\text{QRS}}, \bs{\eta}^{\text{T}}, \bs{\eta}^{\text{P}}$ should approximate only the corresponding waveforms QRS, T, and P. 

\subsection{Bound constraints}
\label{sec:bound_constraints}
The nonlinear model in \eqref{eq:joinmodel} has two variables: the translation $\tau$ and the dilation $\lambda$. These parameters are directly related to medical properties, more specifically the locations and the widths of the QRS, T and P approximations. Analysis of these and other derived parameters (e.g., QT interval) is a very important research field in cardiology, which also provides comprehensive statistics of the standard clinical features of the \ac{ecg}. We apply the results of recent medical studies \cite{RIJNBEEK2014} to derive bound constraints on the values of $\tau$ and $\lambda$.

In order to explain the relationship between dilation $\lambda$ and wave width $|\tilde{I}_0|$, let us consider the first element of the family of rescaled Hermite functions: 
\begin{equation}
	\tilde{\varphi}_0(\lambda,\tau; t) =\pi^{-1/4} \cdot \exp{\left( - (\lambda \cdot (t-\tau))^2/2\right)}\textrm{.}
\label{eq:first_hermite}
\end{equation}
Note that, up to a constant factor, this expression is equal to the probability density function of a normal distribution with mean $\mu=\tau$ and variance $\sigma^2=1/\lambda^2$. Therefore, the well-known three-sigma rule applies here, which means that the spread of $r=1,2,3$ times the standard deviations covers $68\%,\,95\%,\,$ and $99\%$ of the total distribution, respectively. Due to the scaling, this identity roughly applies to all the functions $\tilde{\varphi}_j\,,(j\in\IN)$, and thus also to their linear combinations. As a consequence, the value of the wave approximation is practically zero outside the interval $\tilde{I}_0=[\mu-3\sigma,\mu+3\sigma]$. The wave width is therefore given by $|\tilde{I}_0|=6\sigma=6/\lambda$, and the bounds can be written as follows:
\begin{equation}
	\lambda_{\min}:= \frac{6}{w_{\max}} \leq \lambda \leq \frac{6}{w_{\min}}:=\lambda_{\max}\textrm{,}
\label{eq:bounds}
\end{equation}
where $w_{\min}$ and $w_{\max}$ denote, respectively, the minimum and the maximum widths of the corresponding waveform. Table~\ref{tab:dictpars} summarizes the lower and upper bounds of $\lambda$, for which we chose $w_{\min}$ and $w_{\max}$ according to the clinical statistics of the QRS, T and P waves in \cite{RIJNBEEK2014}.

The medical interpretation of the translation parameter $\tau$ is much easier to describe: It is equal to the center position of the waveform approximation. In the three-sigma rule terminology, $\tau$ simply represents the center of the interval $\tilde{I}_0$. Therefore, timing-based statistics of the QRS, T and P waves directly limit the value of the corresponding translation parameter:
\begin{equation}
	\tau_{\min} \leq \tau \leq \tau_{\max}\textrm{,}
\label{eq:bounds}
\end{equation}
where the upper and lower bounds are defined according to the medical statistics in \cite{RIJNBEEK2014}.

\subsection{Nonlinear constraints}
Additional medical properties can be incorporated into our model by using nonlinear constraints. The most important constraints are the relative positions of the QRS, T and P waves in a heartbeat, which can be formulated as follows:
\begin{gather}
	\label{eq:distinc_support1}
	1 \leq \tau_\text{P}-3/\lambda_\text{P}\textrm{,} \quad \tau_\text{P}+3/\lambda_\text{P} \leq \tau_{\text{QRS}}-3/\lambda_\text{QRS}\textrm{,}\\
	\label{eq:distinc_support2}
	\tau_\text{T}+3/\lambda_\text{T} \leq N\textrm{,} \quad \tau_\text{QRS}+3/\lambda_\text{QRS} \leq \tau_\text{T}-3/\lambda_\text{T}\textrm{.}
\end{gather}
Recall that $\tau \pm 3/\lambda$ is equal to the onset and the offset of the wave approximation provided by the three-sigma rule with $\sigma=1/\lambda$.  Thus, Eq.~\eqref{eq:distinc_support1} means that the onset of the P wave approximation should be greater than or equal to $1$, that is, the first sample index of a beat, while the end of the P wave should be less than the onset of the QRS complex. Similarly, \eqref{eq:distinc_support2} implies that the end of the T wave cannot be greater than the number of samples $N$ in the heartbeat signal, and the end of the QRS approximation should be less than or equal to the onset of the T wave. These conditions guarantee the right ordering of the waveform approximations. Note that the values of the wave approximations are very low outside the corresponding intervals $[\tau - 3/\lambda, \tau+3/\lambda]$. Due to Eqs.~\eqref{eq:distinc_support1} and \eqref{eq:distinc_support2}, these intervals are also distinct, which implies that the dot products of the QRS, T and P components are very small, that is, they are pairwise almost orthogonal (see, e.g., Fig.~\ref{fig:wave_aprx_w_sig}).

A normal cardiac cycle begins with depolarization of the atria (P wave), which is followed by ventricular depolarization (QRS complex). The final phase of the ventricular contraction is the repolarization (T wave), in which the heart returns to its resting state. This means that the recorded electric potential returns to the isoelectric line after the T wave; hence, the cycle starts and ends at this level \cite{ecg_book}. In order to describe this behavior of the ECG, we require the sigmoids of the QRS and of the T wave to have the same coefficients with opposite signs:
\begin{equation}
	\m{c}^{\text{QRS}}_s=-\m{c}^{\text{T}}_s\textrm{.}
\label{eq:equal_sigmoid}
\end{equation}

Additional heuristics can be applied to the parameters of the P wave approximation. It is well known that, in most leads, a normal P wave is associated with a positive deflection away from the isoelectric line \cite{sornmobook}. Therefore, we restrict the coefficient of the first rescaled Hermite function, which is a Gaussian, to non-negative:
\begin{equation}
	0 \leq \m{c}^{\text{P}}_0\textrm{.}
\label{eq:positivity}
\end{equation}
Note that other shapes, such as biphasic and notched P waves can still be represented by higher-order Hermite functions. However, the first Hermite function is forced to be a non-negative component, which supports representation of the most common monophasic morphology. This constraint could be easily switched off in case pathological inverted P waves need to be modelled.

\section{Experiments and Results}

\begin{figure*}[!t]
\centering
\includegraphics[height=18cm]{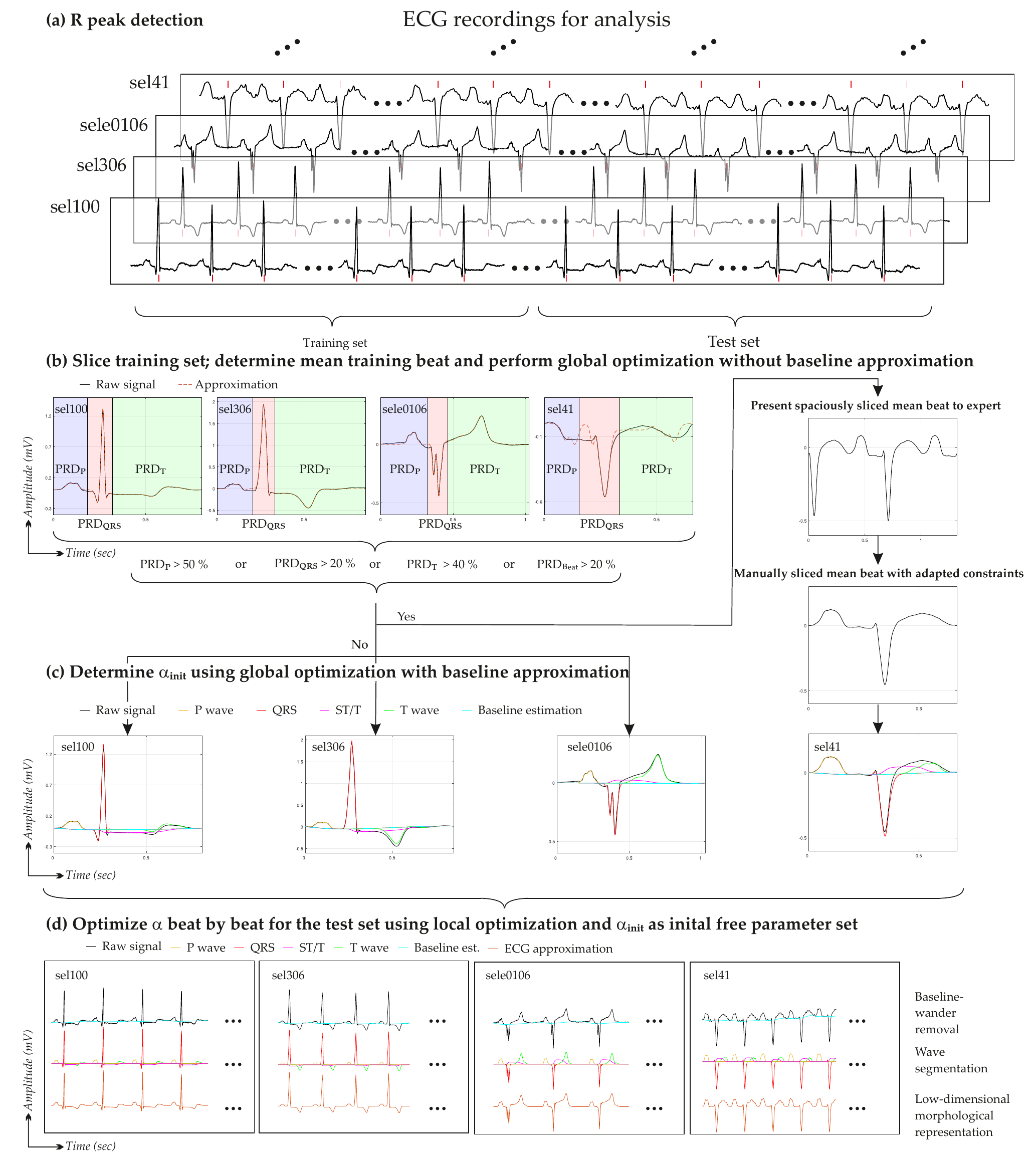}
\caption{Experimental setup for evaluating the effectiveness of our method in terms of baseline elimination and wave segmentation/representation.}
\label{fig:experimental_setup}
\end{figure*}

We used the experimental setup illustrated in Fig.~\ref{fig:experimental_setup} to test the robustness of our algorithm with regard to baseline elimination and wave delineation. In the first step, time locations of the R peaks were determined, which in our case were either already known (from simulated data) or provided by the database used \cite{PhysioNet}. Of course, one could also use computer aided R peak detection, like the well-known Pan-Tompkins algorithm \cite{Pan1985} or more recent methods like \cite{GUPTA2019}. Then we distinguished between a training and a test set, the former one defined to consist of 100 beats preceding the latter one, which represents the \ac{ecg} sequence to be analyzed. As illustrated in Fig.~\ref{fig:experimental_setup}(b) the training set was divided into single beats, where the time instant for slicing was defined to be distance
\begin{equation}
\text{pre}_\text{R} = \frac{\med \left( \text{RR}_\text{train} \right) }{3} 
\end{equation}
preceding each R peak, with $ \text{RR}_\text{train}$ being the sequence of time differences between the R peak locations in the training set. The resulting \ac{ecg} beats were averaged to determine a mean training beat, which was subsequently used to decide whether the chosen setup (i.e., the slicing points and the bound constraints; cf. Sec.~\ref{sec:bound_constraints}) were suited to approximating the corresponding beat. For this reason, translation and dilation parameters were optimized for P-QRS-T by means of a genetic algorithm. On the basis of these optimized parameters, we were able to roughly divide the beat into three segments containing the P wave, the QRS complex and the T wave, respectively (Fig.~\ref{fig:experimental_setup}(b)). We then calculated the \ac{prd} between the raw signal $\mathbf{f}$ and its approximation $\mathbf{\hat{f}}$ ($\overline{\mathbf{f}}$ represents the beat average), 
\begin{equation*}
\text{PRD}_\text{Beat} = \frac{||\mathbf{f} - \mathbf{\hat{f}}||_2}{||\mathbf{f} - \overline{\mathbf{f}}||_2} \cdot 100   
\end{equation*}
for the whole beat and for the respective segments
\begin{align*}
\text{PRD}_\text{P} &= \frac{||\mathbf{f}_\text{P} - \mathbf{\hat{f}_\text{P}}||_2}{||\mathbf{f}_\text{P} - \overline{\mathbf{f}}_\text{P}||_2} \cdot 100,\,\, \text{PRD}_\text{QRS} = \frac{||\mathbf{f}_\text{QRS} - \mathbf{\hat{f}_\text{QRS}}||_2}{||\mathbf{f}_\text{QRS} - \overline{\mathbf{f}}_\text{QRS}||_2} \cdot 100, \\
\text{PRD}_\text{T} &= \frac{||\mathbf{f}_\text{T} - \mathbf{\hat{f}_\text{T}}||_2}{||\mathbf{f}_\text{T} - \overline{\mathbf{f}}_\text{T}||_2} \cdot 100.
\end{align*}
We compared these values to experimentally determined thresholds (defined in Fig.~\ref{fig:experimental_setup}(b)), assuming that the approximation of the whole beat or single segments result in suitable slicing and bound constraints. Certainly, this should apply to the majority of \ac{ecg} recordings analyzed, as the default setup (i.e., default slicing points and bound constraints) is based on \cite{RIJNBEEK2014}. However, if the slicing is incorrect, or in the case of very atypical beat morphologies, a higher \ac{prd} of one or more segments or even of the whole beat indicates that manual annotation of the mean beat is necessary. This leads to an adapted slicing while implicitly redefining the bound constraints, which -- due to an incorrect slicing and an unnaturally long P wave -- was the case for recording sel41 in our illustration (Fig.~\ref{fig:experimental_setup}(b)). Note that in this step we did not optimize for possible \ac{blw}, since incorrect slicing in particular may be compensated for by the splines, and, falsify the decision.

Once the mean beat and the related waves were determined, we optimized for an initial $\bs{\alpha}_\text{init}$ per recording, this time taking a small potential \ac{blw} into account (Fig.~\ref{fig:experimental_setup}~(c)). Based on these individual, pre-optimized parameters, the single \ac{ecg} beats of the test set were then approximated by adjusting the free parameters $\bs\alpha_i, (i = 1,\ldots,\mathcal{I})$ by means of local nonlinear optimization (Fig.~\ref{fig:experimental_setup}~(d)). 

\subsection{Baseline wander removal} \label{sec:baseline}
\begin{figure}[!t]
\centering
  \subfigure[Clean ECG superimposed with simulated baseline noise.]{
  \includegraphics[scale=0.3]{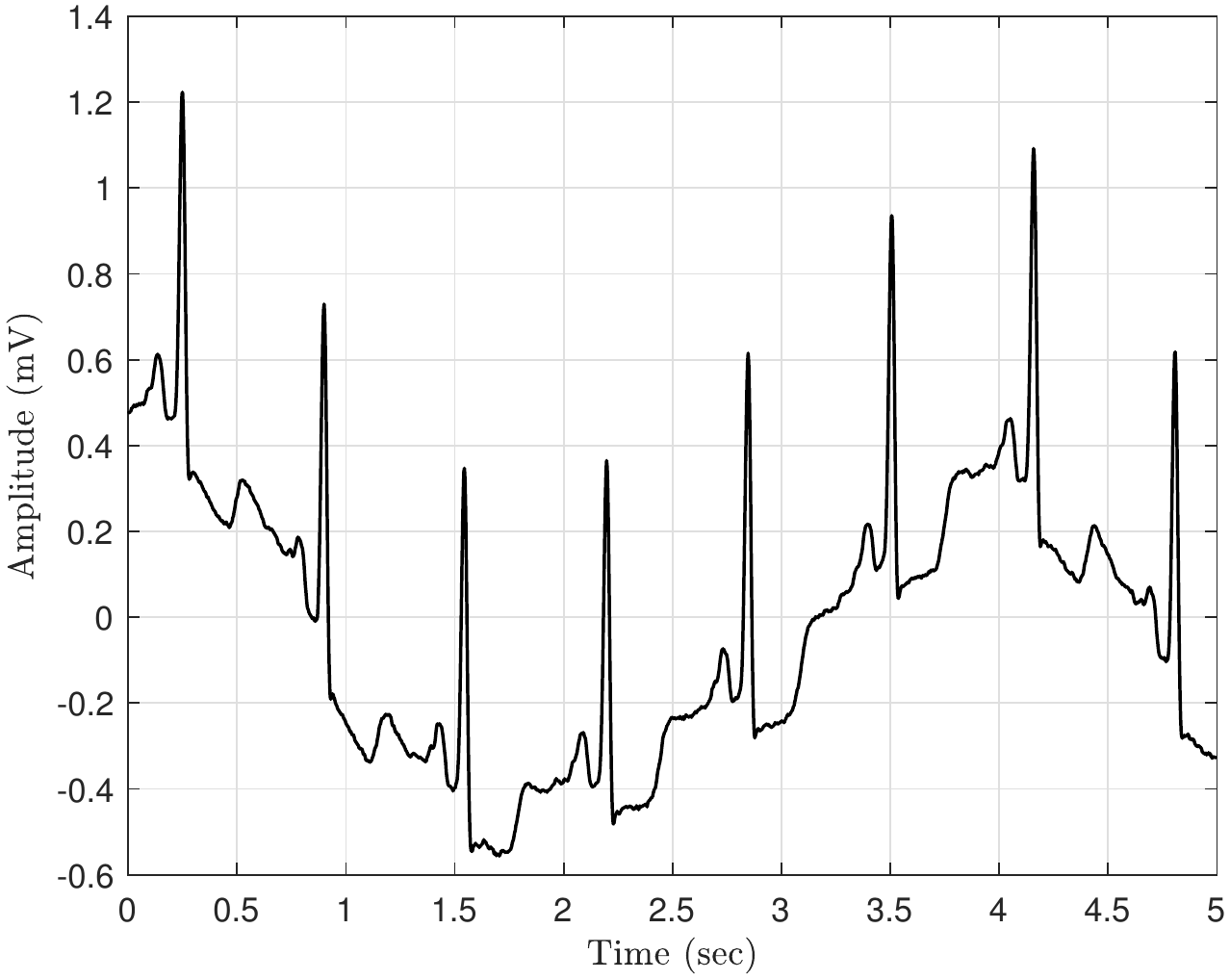}
  \label{fig:bl_noisy_iir_demo}
  } 
  \subfigure[Clean ECG and denoised ECG.]{
  \includegraphics[scale=0.3]{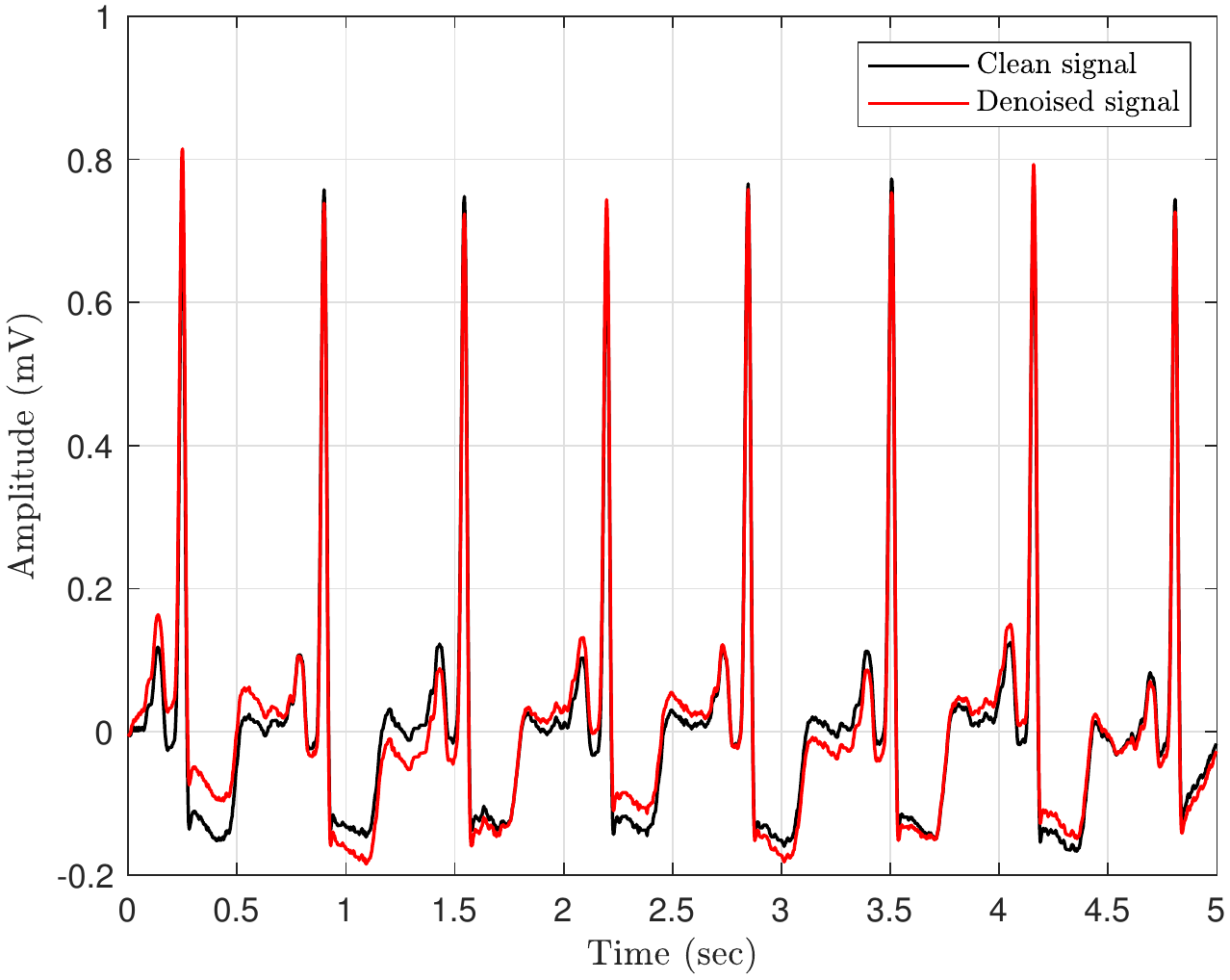}
  \label{fig:bl_remove_iir_demo}
  } 
\caption{Demonstrating the effect of baseline elimination on the ECG (ST segment and T wave are altered in this case).}
\label{fig:bl_demo_iir}
\end{figure}

In our first experiment, we assessed the effectiveness of our method in removing \ac{blw} while preserving important diagnostic information. \ac{blw} is a low-frequency artifact in the \ac{ecg} that can have several sources, such as breathing and patient or electrode movement. Since it is such a common phenomenon, several state-of-the-art techniques for \ac{blw} removal have been developed, such as spline interpolation, wavelet cancellation, and median filtering. A major issue in \ac{blw} removal is distortion of important diagnostic information, for instance, changes in the ST segment. Applying the denoising algorithm may alter the ST segment, resulting in an ST level that differs from the original (i.e., clean) one, which could lead to a different diagnosis in terms of \ac{ste}/\ac{std} (Fig.~\ref{fig:bl_demo_iir}). The reason for this is that most state-of-the-art algorithms focus on the removal of the baseline without taking morphological distortion into account. Our proposed method, in contrast, considers these two tasks simultaneously, which makes it a nonlinear baseline-removing and morphological feature preserving filter for \ac{ecg}.

Evaluating algorithms in terms of their ability to retain the correct diagnostic information, Lenis et al. have recently published a simulation study that compares five state-of-the-art filtering techniques for baseline removal \cite{Lenis2017}. In particular, they focused on preserving ST segment diagnostic information in the ischemic heart. Since they allowed us to use their simulated \ac{ecg} beats \cite{Loewe2011,Loewe2015}, we were able to carry out a similar study, comparing our work to various state-of-the-art baseline elimination algorithms. Hence, we provide a brief review of their methodology for data generation and algorithm evaluation before presenting the results of our algorithm compared to those of selected \ac{blw} removal techniques. 

\begin{figure*}[!ht]
\centering
\subfigure[Example ECG with ST depression.]{
\includegraphics[width=5.5cm]{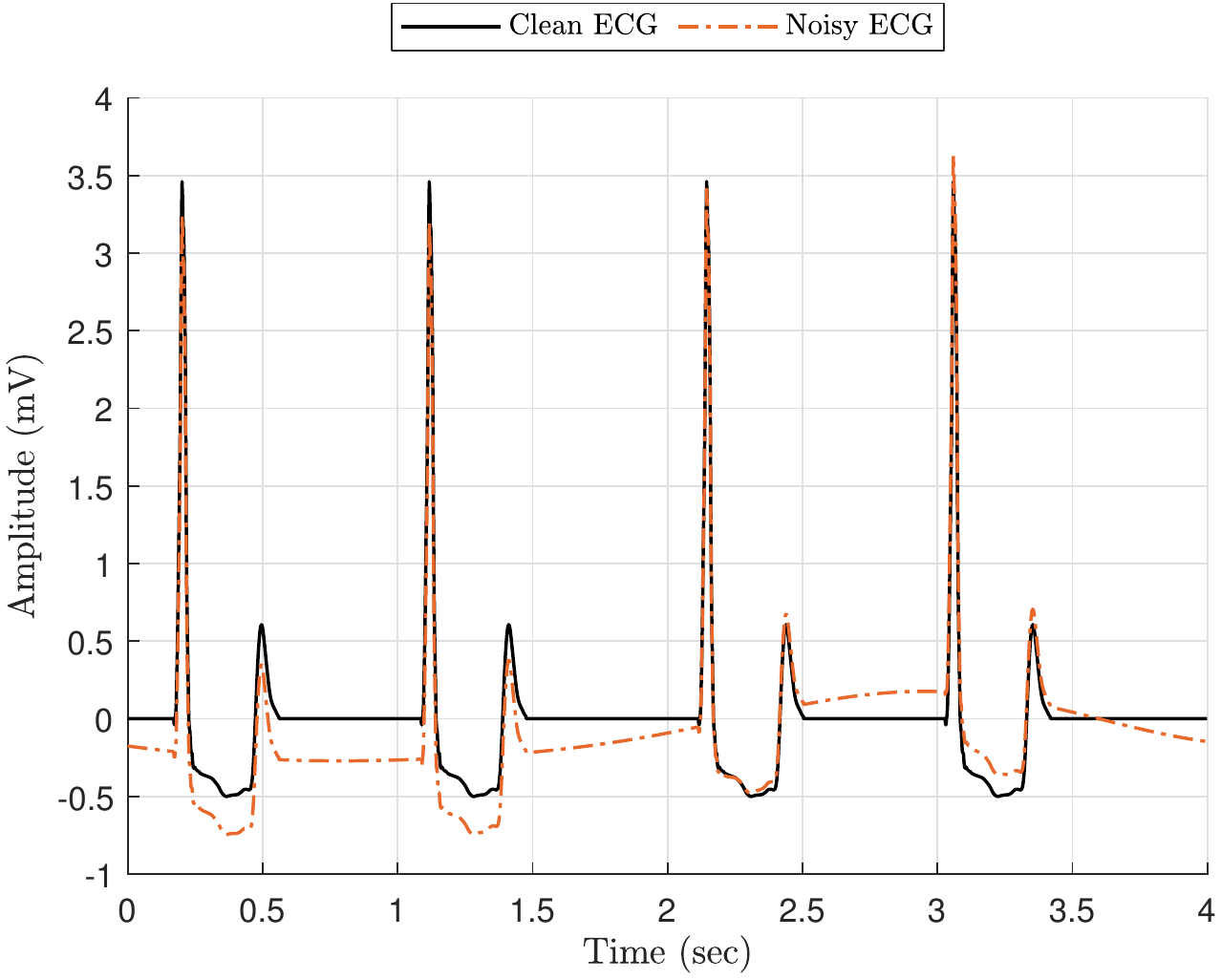}
\label{fig:ecg_clean_10dB}
}
\subfigure[Example ECG  with no obvious pathology.]{
\includegraphics[width=5.5cm]{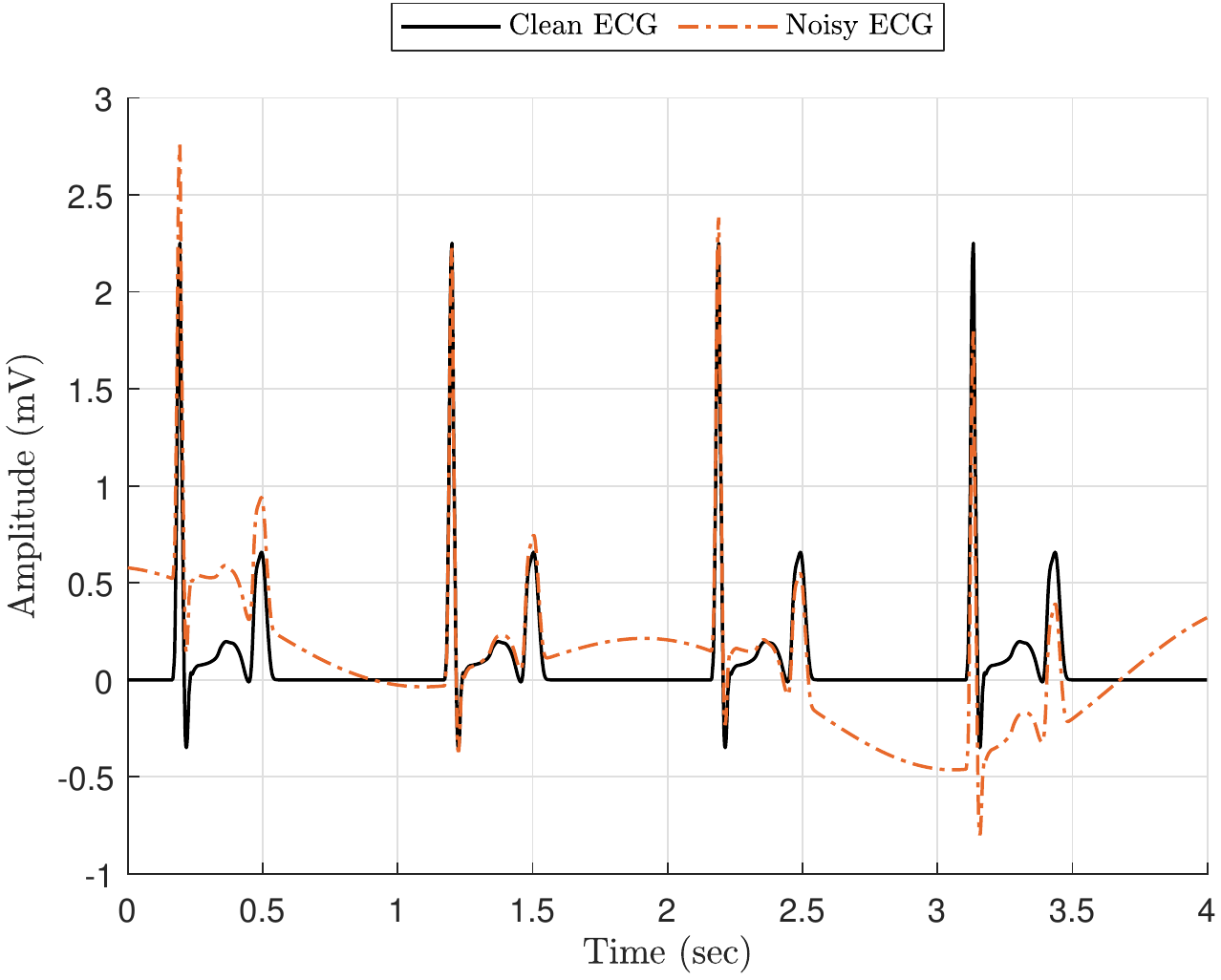}
\label{fig:ecg_clean_0dB}
}
\subfigure[Example ECG with ST elevation.]{
\includegraphics[width=5.5cm]{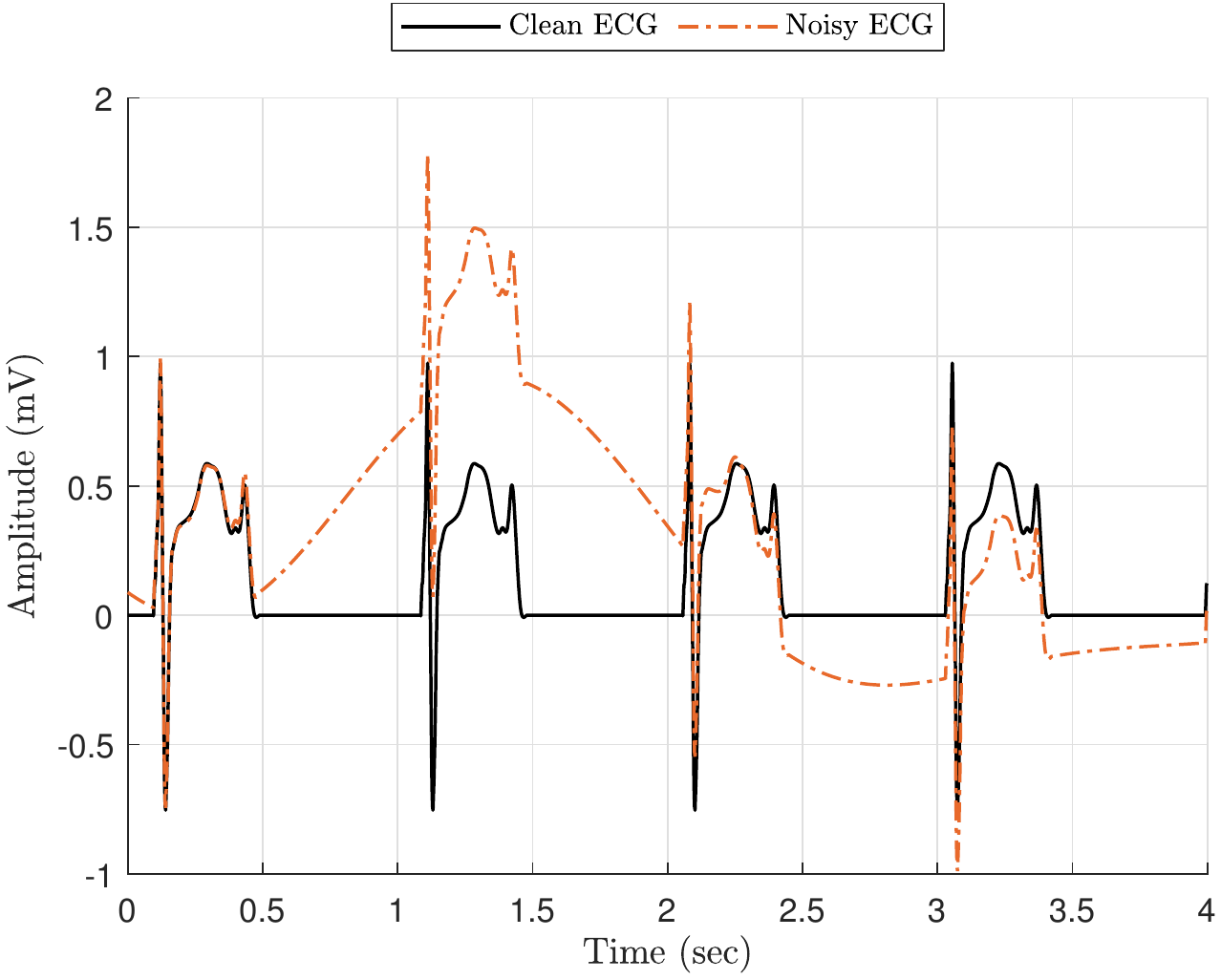}
\label{fig:ecg_clean_m10dB}
}

\subfigure[Clean and denoised ECG contaminated by slight baseline noise, shown in (a).]{
\includegraphics[width=5.5cm]{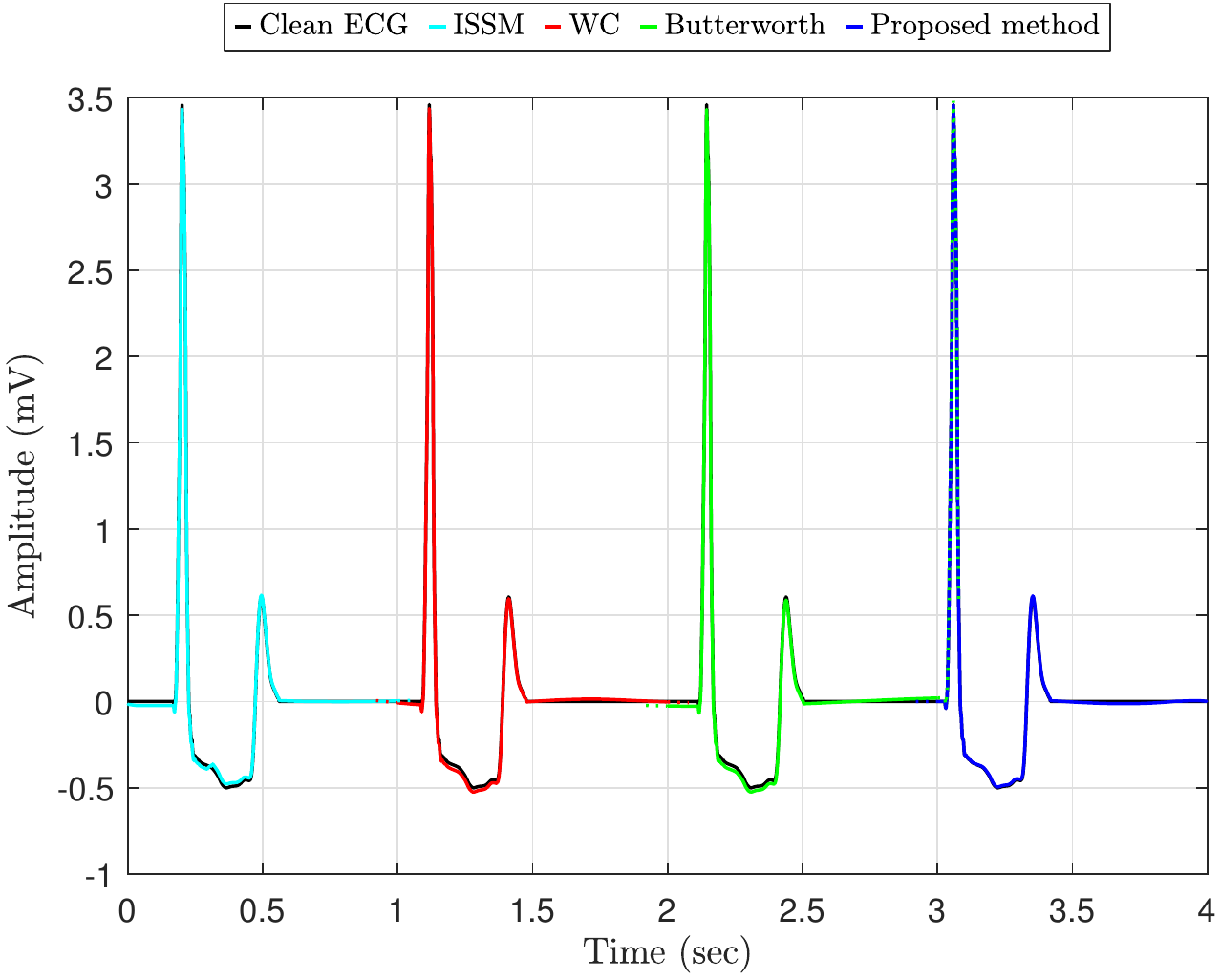}
\label{fig:ecg_denoised_10dB}
}
\subfigure[Clean and denoised ECG contaminated by moderated baseline noise, shown in (b).]{
\includegraphics[width=5.5cm]{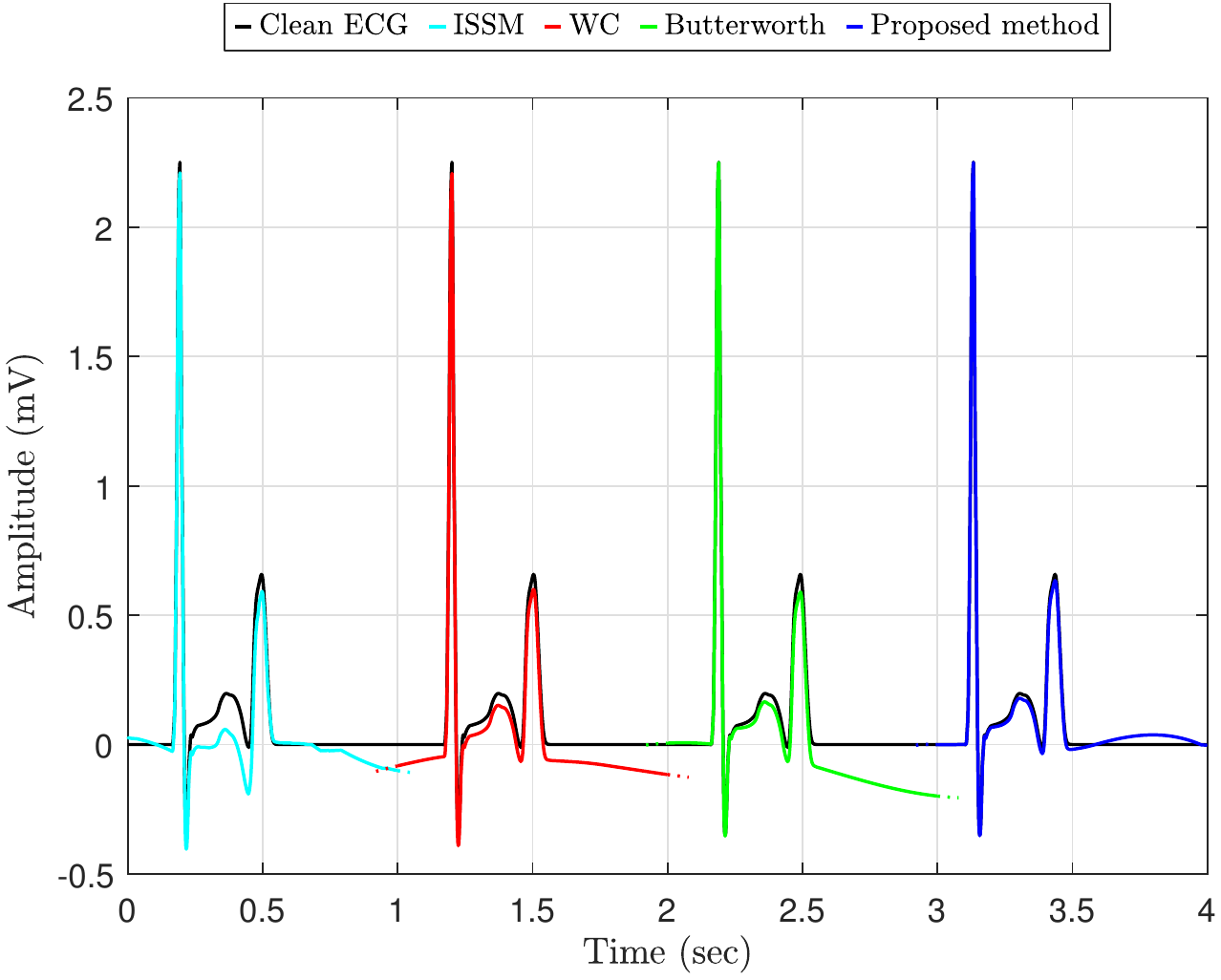}
\label{fig:ecg_denoised_0dB}
}
\subfigure[Clean and denoised ECG contaminated by strong baseline noise, shown in (c).]{
\includegraphics[width=5.5cm]{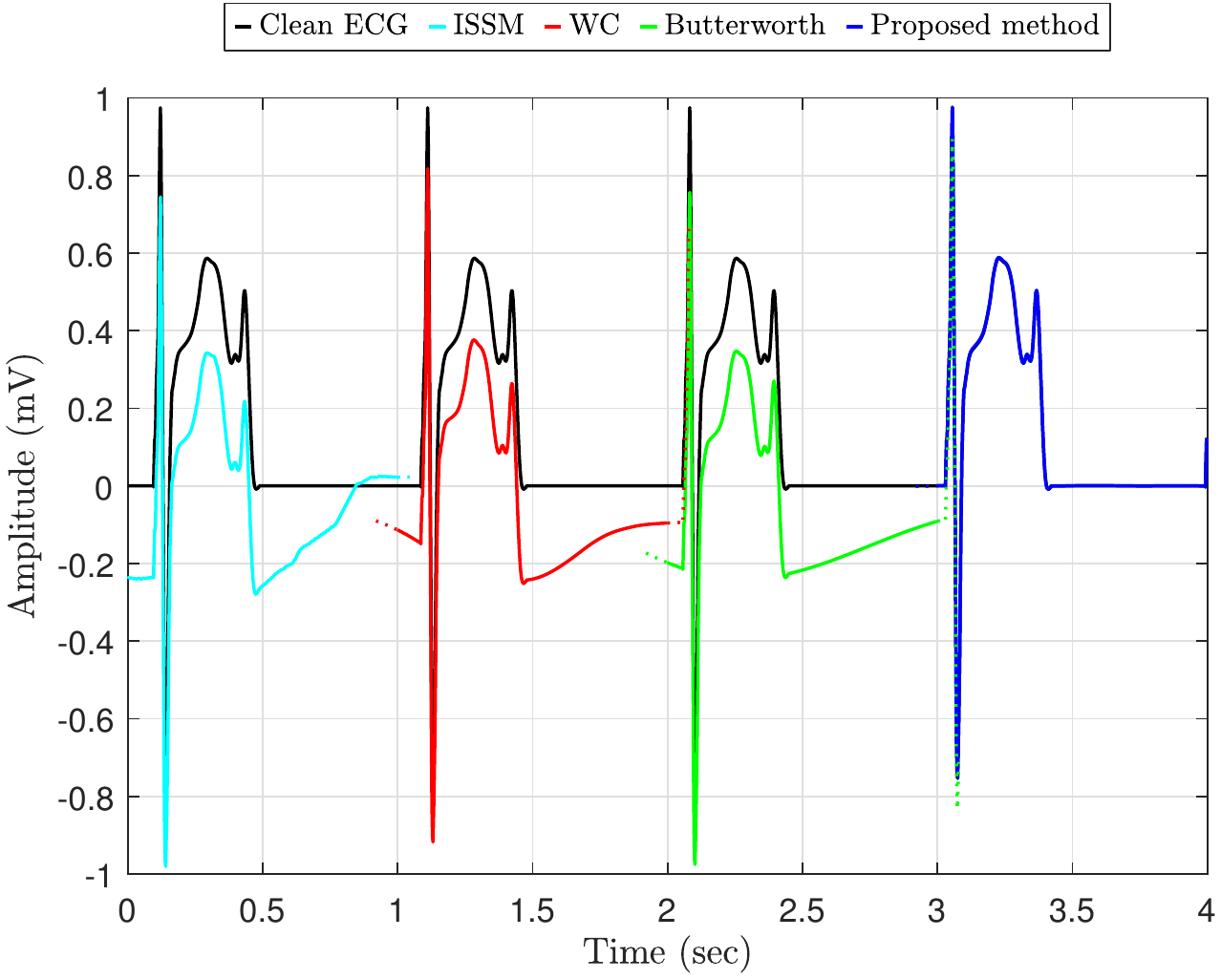}
\label{fig:ecg_denoised_m10dB}
} 
\caption{Effectiveness of four selected \ac{blw} removal techniques for three example \ac{ecg} recordings contaminated with 3 different types of baseline noise (a)-(c). Clean and denoised \ac{ecg} recordings (d)-(f); for better visibility only one beat is shown per denoising method.}
\label{fig_intro}
\end{figure*}

\subsubsection*{ECG and noise generation}
We reused the dataset originally proposed by Loewe et al. for optimal electrode placement for the ischemic heart\cite{Loewe2011, Loewe2015}. This dataset contains simulated surface \ac{ecg}s from 3 different subjects with several degrees of ischemia, amounting to $3 \times 255 = 765$ electrophysiological setups in total. Each of these setups is represented by a 12-lead \ac{ecg}, where one lead consists of a single beat (QRS complex, ST segment, and T wave) sampled at $f_s = 500\,$Hz. The beats were resampled  to $f_s = 512\,$Hz such that wavelet-based filtering, as done in \cite{Lenis2017}, could be carried out for comparison. In order to obtain recordings of reasonable length ($250 \,$s), we quasiperiodically extended the single leads as suggested in \cite{Lenis2017}. We added variable RR intervals following a Gaussian distribution with $\mu=1\,$s and $\sigma=50\,$ms. Thus, we obtained $765$ clean $12$-lead \ac{ecg} recordings with specific diagnostic information given by the ST segment (Fig.~\ref{fig:ecg_clean_10dB}-\ref{fig:ecg_clean_m10dB}). These clean recordings were subsequently contaminated with baseline noise, defined as
\begin{equation}
n_{\text{BL}}(t) = C \cdot \sum\limits_{k=0}^{K} a_k \cdot \cos\left( 2 \pi \cdot k \cdot \Delta f \cdot t + \theta_k \right),
\end{equation}
where, in accordance with \cite{Lenis2017}, $\Delta f$ and $K$ were set to $0.01\,$Hz and $50$, respectively, while $a_k$ and $\theta_k$ were defined to be random numbers drawn from a uniform distribution within the intervals $[0,1]$ and $[0,2\pi)$, respectively. The weighting factor $C$ determines the total power of the baseline signal; varying this factor allows experiments to be carried out for different \ac{snrs}. The \ac{snrs} were defined to be $-10\,$dB, $0\,$dB, and $10\,$dB to obtain realistic \ac{ecg} recordings with slight, moderate, and strong baseline noise (Fig.~\ref{fig:ecg_clean_10dB}-\ref{fig:ecg_clean_m10dB}). Consequently, a total of $27,540$ (=$765 \times 12 \times 3$) \ac{ecg} traces was produced, which allowed us to compare our method to various other denoising algorithms on a big dataset.

\subsubsection*{Denoising algorithms}
Most of the baseline removal algorithms we compared our work to, were reimplemented and reviewed in \cite{Romero2018}. In particular, we considered FIR highpass filtering \cite{Alste1985}, IIR highpass filtering \cite{Pottalla1990}, cubic splines \cite{Meyer1977}, adaptive filtering \cite{Laguna1992}, moving-average filtering \cite{Canan1998}, median filtering \cite{Chouhan2007}, and wavelet-based baseline cancellation \cite{Lenis2017}. These algorithms are regularly cited in the literature and therefore provide a good benchmark for judging our method, based on the following quality criteria.

\subsubsection*{Quality criteria}
Morphological distortion and removal of diagnostic information is well described  by four measures: the SNR,  the correlation coefficient, the so-called $l\_$operator, and the deviation of the ST change \cite{Lenis2017}. 

First, the SNR is a well-known measure for evaluating the denoising performance of an algorithm. Since we were using synthetic data for this experiment, the clean signal is known and consequently we are able to determine exact output SNR values to compare various denoising algorithms.

Second, the correlation coefficient mainly quantifies the morphological matching of the original and the reconstructed signal, independent of scaling and offsetting of the signals. Taking the clean ECG recordings $x(t)$ and the denoised ones $\hat{x}(t)$, the correlation coefficient is defined as
\begin{equation}
\rho\{x(t), \hat{x}(t)\} = \frac{\EX{\left\{ \left(x(t) - \mu_x \right) \left( \hat{x}(t) - \mu_{\hat{x}} \right) \right\} }}{\sigma_x \sigma_{\hat{x}}},
\label{eq:cc}
\end{equation}
where $\EX\{\}$ denotes the expected value, $\mu_x$ and $\mu_{\hat{x}}$ the expected values of the clean and the denoised \ac{ecg} recordings, respectively, and $\sigma_x$ and $\sigma_{\hat{x}}$ the standard deviations of the clean and the denoised \ac{ecg} recordings.

Third, in order to take offsetting and scaling into account, in \cite{Lenis2014} the so-called $l\_$operator, defined as
\begin{equation}
l\_\text{operator}\{x(t), \hat{x}(t)\} = 1 - \frac{\EX{\left\{ x(t) - \hat{x}(t) \right\} } }{ \EX{\left\{ x^2(t) \right\} }  \EX{\left\{ \hat{x}^2(t) \right\} }},
\label{eq:lop}
\end{equation}
was introduced. It is based on the Euclidean distance between the two signals and gives a value between $-1$ and $1$, where $-1$ corresponds to a complete mismatch and $1$ stands for a perfect alignment. 

Fourth, in this experiment the quantification of the ST change is probably the most important parameter in the context of diagnostic information distortion. Usually, a specific time instant within the ST segment, the so-called J point, is evaluated for diagnosis. There exist clear recommendations for ST segment interpretations depending on specific pre-defined thresholds \cite{Wagner2009}, where deviations by as little as $50\,\mu V$ may be of medical interest. However, since automated determination of the J point is relatively difficult, Loewe et al. introduced the so-called K point (KP), which is also located within the ST segment and provides a measure that is equivalent to the J point from a diagnostic point of view \cite{Loewe2015}. The value of this characteristic time instant can be calculated automatically by first determining the envelope of the twelve lead \ac{ecg}, denoted as $x_i(t),\,\, i = 1,\ldots,12$, and subsequently taking the minimum value of the ST segment:
\begin{equation}
\text{KP} = \min_{t_{\text{ST}}} \left( \max_i |x_i(t_{\text{ST}})| \right), \qquad (i = 1, \ldots, 12), 
\label{eq:kp}
\end{equation}
where $t_{\text{ST}}$ stands for the length of the time interval of the ST segment. Finally, the deviation of the KP is defined as
\begin{equation}
\Delta \text{KP} = \text{KP}_\text{filtered} - \text{KP}_\text{clean}. 
\label{eq:kp_dev}
\end{equation}

\subsubsection*{Method evaluation}
A total of $27,540$ \ac{ecg} traces was generated, each of $250\,s$ length, which corresponds to approximately $250$ beats. However, for method comparison, we considered only the \ac{ecg} traces between beats $100$ and $200$, obtaining a test set of $98$ beats or approximately $100\,s$. This was done for two reasons: First, this way we were sure to avoid edge effects for several filtering methods, and, second, the first $100$ beats could be used to determine an initial parameter set $\bs\alpha_\text{init}$ for our method, as shown in Fig.~\ref{fig:experimental_setup}. 

The noisy \ac{ecg} recordings were then denoised by several filtering methods and subsequently quality criteria were calculated. In order to assess the performance of each method, median and \ac{iqr} were calculated for  the SNR, the correlation coefficient, the $l\_$operator, and the KP deviation. For the first three measures, a method is considered superior if its median is statistically significantly higher than those of the other methods. Clearly, the KP deviation should be closest to zero for a method to perform well. Statistical significance was tested by using the Wilcoxon signed rank test and a significance level of  $5\,\%$ \cite{Lenis2017}. Additionally, we compared the \ac{iqr}s of all methods.  

\subsubsection*{Results}

\begin{table*}[!ht]
\caption{Performance comparison based on SNR, correlation coefficient, $l\_$operator, and KP deviation between several baseline elimination algorithms and the proposed approach. Results are given as median ($25^\text{th}$ percentile, $75^\text{th}$ percentile).}
\centering
\scalebox{0.9} { { \renewcommand{\arraystretch}{1.3} 
\begin{tabular}{|c c c c c|}
\hline
Method & SNR improvement (dB) & $\rho$ & $l\_$operator & KP dev. (mV) \\ \hline
No filter \cite{Alste1985} & - & $0.7026 ~(0.3036,0.9503)$&$ 0.6666 ~(0.1738,0.9513)$&$ -0.0009 ~(-0.1943,0.1912)$ \\
FIR \cite{Alste1985} & $10.6763 ~(6.7094,12.7492)$ & $0.9753 ~(0.8674,0.9963)$&$ 0.9471 ~(0.8233,0.9746)$&$  0.0202 ~(-0.0595,0.1026)$ \\
Mov. avg. \cite{Canan1998} & $12.4538 ~( 7.1013,14.5636)$ & $0.9844~(0.8862, 0.9976 )$&$ 0.9547 ~(0.8613 , 0.9799) $&$  0.0205 ~(-0.0527, 0.0945)$ \\ 
IIR \cite{Pottalla1990} & $12.9428 ~(7.2486,15.3180)$ & $0.9873~(0.9123, 0.9982)$&$ 0.9565  ~ (0.8812,0.9818) $&$  0.0207  ~ (-0.0509,0.0927)$ \\                           
ISSM \cite{Chouhan2007} & $15.1653  ~(11.8574 ,17.7468 )$  & $0.9876~(0.9443 , 0.9957)$&$ 0.9848 ~(0.9367,0.9945)$&$  0.0064 ~(-0.0379,0.0544)$ \\
Spline \cite{Meyer1977} & $14.1993 ~(12.3134,26.0353)$ & $0.9894~(0.9600 ,0.9984 )$&$ 0.9896   ~(0.9543  ,0.9984 ) $&$  0.0004 ~(-0.0113, 0.0173)$ \\
Wav. canc. \cite{Lenis2017} & $11.6437 ~(3.8148,19.8817)$  & $0.9841 ~(0.9677 ,0.9903 )$&$ 0.9627  ~(0.9402 , 0.9821 ) $&$  0.0259 ~-0.0634 ,0.1075)$ \\                                                   
Prop. approach & $27.6623 ~(26.6066,28.6637)$ & $0.9991 ~(0.9932 ,0.9999 )$&$ 0.9991 ~(0.9935,0.9999 ) $&$  0.0000~(-0.0079,0.0079)$ \\ \hline
\end{tabular}
}}
\label{tbl:bl_res}
\end{table*} 
 
Table~\ref{tbl:bl_res} illustrates that our approach outperformed the other algorithms in all four quality criteria. Concerning the SNR, our method showed an average improvement of more than $27$~dB, which is clearly above the SNR improvements of the competing algorithms. Also, for the correlation coefficient and the $l\_$operator, our method had the highest medians, and the median of the KP deviation was closer to zero than for any other algorithm. The Wilcoxon signed rank test showed a statistically significant difference between the results of our method and those of the six \ac{blw} removal techniques. Furthermore, the scatter quantified by the \ac{iqr} was the smallest in our case for all quality criteria, which indicates that our method is the most robust in different scenarios of baseline noise. This is especially noticeable for the SNR improvement, showing an \ac{iqr} of only $2.1$~dB for the proposed work, clearly outperforming the other methods. Fig.~\ref{fig:ecg_denoised_10dB}-\ref{fig:ecg_denoised_m10dB} illustrate these observations in a more intuitive way (we only present 4 example denoising algorithms for better visibility). The selected denoising algorithms reduce the baseline influence effectively, but in some cases diagnostic information (i.e., the level of the ST segment) is corrupted. The last beats of Fig.~\ref{fig:ecg_denoised_10dB}-\ref{fig:ecg_denoised_m10dB} show that our method retains the diagnostic information almost perfectly, confirming that it performed better than the other algorithms.

\subsection{Wave segmentation} \label{sec:wave_segmentation}
Wave segmentation (also: wave delineation) is usually defined as determining onset, peak, and end of the waves (i.e., P-QRS-T) and remains one of the major challenges in \ac{ecg} signal processing. Not only is automated wave delineation difficult to accomplish itself, the lack of a gold-standard evaluation methodology also makes this task extremely demanding. Nevertheless, a reliable delineation algorithm is essential in clinical applications, since the analysis of \ac{ecg} time domain parameters (e.g., the QT interval) often plays an important role in determining a patient's treatment, for instance, whether lifelong medication is indicated \cite{Bennet2014}. In our case, an adequate wave segmentation is crucial to describing the morphological development of individual waves. Consequently, we evaluated the performance of our method using the Physionet \ac{qtdb}  \cite{PhysioNet, Laguna1997} to demonstrate that it delivers results that are comparable to those of the latest ECG beat delineation algorithms. Despite several limitations \cite{Gonzalez2017}, the \ac{qtdb} remains one of the best options for testing the robustness of a delineation algorithm; furthermore, it has been used by several research groups to compare delineation results. In the next three paragraphs, we thus describe the delineation method, the \ac{qtdb}, and our results. 

\subsubsection*{Delineation}
\begin{figure}[!t]
\centering
\includegraphics[width=8cm]{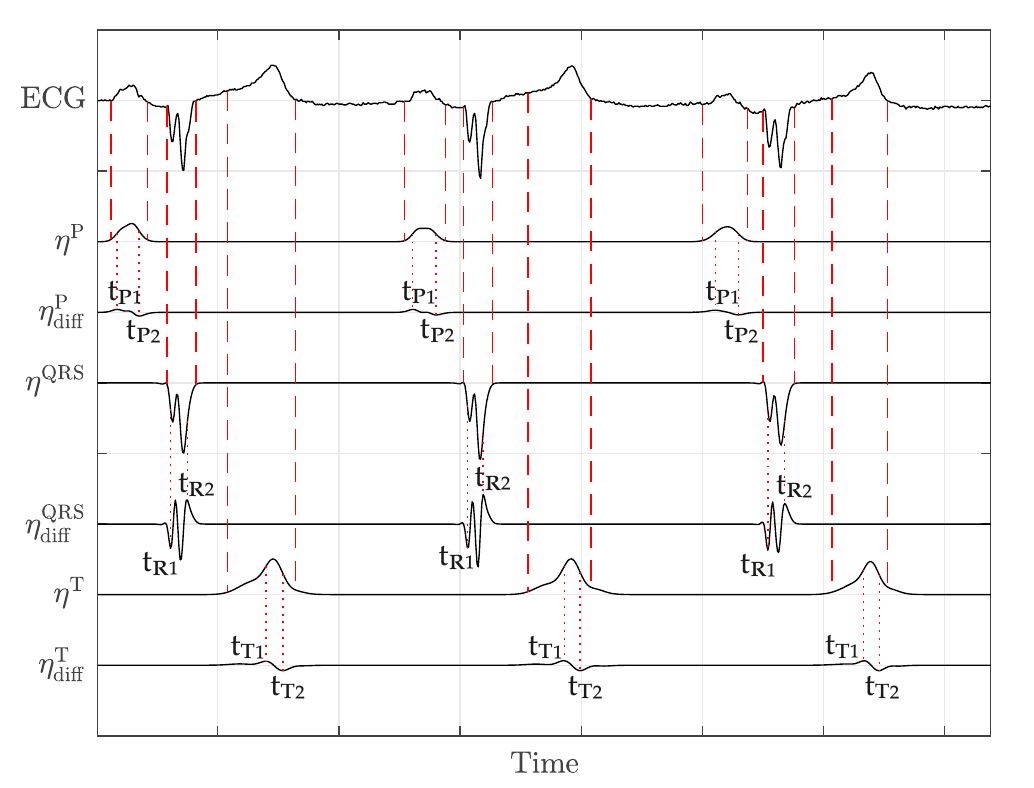}
\caption{Demonstration of the \ac{ecg} delineation method.}
\label{fig:del_demo}
\end{figure}

As illustrated in Fig.~\ref{fig:experimental_setup}(d), the basic segmentation into the main components is already given by carrying out the local optimization for all the beats of interest (i.e., $\bs{\eta}^{\text{P}}$, $\bs{\eta}^{\text{QRS}}$ and $\bs{\eta}^{\text{T}}$). The locations of the peaks can be easily determined by finding the maximum values of the approximations. In order to determine onset and end of the wave approximations, we used the information provided by their derivatives (Fig.~\ref{fig:del_demo}, inspired by \cite{Laguna1994}). The following explanations apply to each beat individually. Note that the approximation represents the \ac{ecg} in an analytic form, allowing the derivatives to be calculated analytically. Furthermore, except for wave-depending thresholds, the methodology is the same for all three components (P-QRS-T). We therefore describe it only once for the QRS complex and list the thresholds used (which were determined experimentally) in Table~\ref{tbl:th_del}.

Starting with the derivative of a single QRS complex $\bs{\eta}^{\text{QRS}'}$, we eliminated all maxima and minima lower/larger than $\max\left(\bs{\eta}^{\text{QRS}'}\right) / 20$ and $\min\left(\bs{\eta}^{\text{QRS}'}\right) / 20$, respectively. Subsequently, to determine the onset, we took the leftmost of the remaining maxima / minima, located at $t_\text{R1}$ (Fig.~\ref{fig:del_demo}), and defined two potential candidates left to $t_\text{R1}$ for the onset:
\begin{itemize}
\item Candidate 1: $|\bs{\eta}^{\text{QRS}'}|$ falls below the threshold $$ \text{th}_\text{on}^\text{QRS} = 0.05 \cdot |\bs{\eta}^{\text{QRS}'}(t_\text{R1})|.$$
\item Candidate 2: A local maximum / minimum is detected left to $t_\text{R1}$ for a negative / positive $\bs{\eta}^{\text{QRS}'}(t_\text{R1})$. 
\end{itemize}
We then defined the candidate closer to $t_\text{R1}$ to be the onset. A very similar procedure was carried out to determine the end of the QRS complex, where we searched to the right of the last significant maximum / minimum $(t_\text{R2})$ for defining the two candidates as described above. Additionally, a slightly altered threshold was used, as illustrated in Table~\ref{tbl:th_del}. Finally, we obtained onset, peak, and end of three waves (P-QRS-T) for all beats under investigation, which were compared to expert annotations given for the \ac{qtdb} for evaluation. 

\begin{table}
\center
\caption{Thresholds used for \ac{ecg} wave delineation.}
\label{tbl:th_del}
\scalebox{0.8} { { \renewcommand{\arraystretch}{1.3} 
\begin{tabular}{|c | c c c|}
\hline
 & Significant max/min & $\text{th}_\text{on}$ & $\text{th}_\text{end}$ \\ \hline
\multirow{2}{*}{P} & $\max\left(\bs{\eta}^{\text{P}'}\right) / 2$ & \multirow{2}{*}{$0.25 \cdot |\bs{\eta}^{\text{P}'}(t_\text{P1})|$} & \multirow{2}{*}{$ 0.4 \cdot |\bs{\eta}^{\text{P}'}(t_\text{P2})|$}  \\
 &  $\min\left(\bs{\eta}^{\text{P}'}\right) / 2$ & & \\
 \multirow{2}{*}{QRS} & $\max\left(\bs{\eta}^{\text{QRS}'}\right) / 20$ & \multirow{2}{*}{$0.05 \cdot |\bs{\eta}^{\text{QRS}'}(t_\text{R1})|$} & \multirow{2}{*}{$ 0.07 \cdot |\bs{\eta}^{\text{QRS}'}(t_\text{R2})|$}  \\
 &  $\min\left(\bs{\eta}^{\text{QRS}'}\right) / 20$ & & \\
 \multirow{2}{*}{T} & $\max\left(\bs{\eta}^{\text{T}'}\right) / 2$ & \multirow{2}{*}{$0.25 \cdot |\bs{\eta}^{\text{T}'}(t_\text{T1})|$} & \multirow{2}{*}{$ 0.4 \cdot |\bs{\eta}^{\text{T}'}(t_\text{T2})|$}  \\
  &  $\min\left(\bs{\eta}^{\text{T}'}\right) / 2$ & & \\
 
 \hline
 
\end{tabular}
}}
\end{table}

\subsubsection*{The QT database}

There are mainly three reasons why we chose the \ac{qtdb} for assessing our algorithm's delineation ability. First, it contains expert annotations for more than 3000 beats of 105 2-channel \ac{ecg} recordings, which provides a large variety of beat morphologies and allows a method's robustness  against pathological / atypical wave forms to be estimated. Second, it has been used extensively by other research groups, and therefore we can easily estimate the general delineation quality by comparing its results to those provided in the literature \cite{Gonzalez2017}. Lastly, our previous approach, where we used adaptive Hermite functions without spline interpolation and sigmoidal functions for \ac{ecg} wave delineation \cite{Kovacs2017}, was also evaluated on this database. Consequently, we are able to directly illustrate the improvement we achieved in terms of delineation. 

The strategy for comparing single-channel delineation algorithms using the \ac{qtdb}, which we also followed in this work, was originally proposed in \cite{Martinez2004}. First, we determined the time differences between expert and algorithm annotations for every \ac{ecg} characteristic point labelled by the expert. In order to address the issue that the expert annotated the recordings by looking at both \ac{ecg} channels at the same time while the algorithm performs single-lead delineation, we followed the recommendations given in \cite{Martinez2004}. This means that, for every \ac{ecg} characteristic point, we chose the channel with the smaller error between expert and algorithm annotation. Subsequently, for every recording a bias $\mu_e$ and a standard deviation $\sigma_e$ of the respective deviations were calculated. Additionally, sensitivity of the wave detection itself was determined to quantify the number of waves that were detected by the expert but not by the algorithm.

This evaluation strategy adequately estimates the overall-performance of the delineation algorithm, but it lacks a means of quantifying the percentages of recordings in which $\mu_e$ and $\sigma_e$ are within acceptable limits and those in which they are not. In order to achieve this, we split the dataset into four groups by defining acceptable tolerances for $\mu_e$ and $\sigma_e$ per \ac{ecg} characteristic point, as illustrated on the left side in Table~\ref{tbl:split_ds} \cite{Jane1997, Kovacs2017}. Group I holds all recordings with low bias and standard deviation, which is clearly the targeted group. In the case of $\text{P}_\text{on}$, for instance, this would mean that the bias must be smaller than $25\,\text{ms}$, while the $\sigma_e$ should be smaller than $30\,\text{ms}$ for a recording to be assigned to group I. For group II the difference between expert and algorithm annotations is greater, which does not automatically mean poor performance, because in some cases even experts disagree on the true value of the characteristic point \cite{Gonzalez2017}. Groups III and IV have a high standard deviation, which indicates varying algorithm (or expert) annotations for similar beats within a recording and is, of course, not desired at all. The main reason for this is usually low general signal quality, which cannot be handled well by the algorithm and sometimes not even by the expert. 
\subsubsection*{Results}
Using the general methodology shown in Fig.~\ref{fig:experimental_setup}, we obtained 81 recordings for which the global mean beat was  approximated in an adequate manner. Consequently, these recordings could be processed using the standard \ac{ecg} beat-slicing method and the standard constraints for $\lambda$ and $\tau$, and manual mean beat annotations were not required. Of course, this also means that, prior to automated delineation by the algorithm, the mean beat of 24 recordings had to be labelled manually, which is a high number given the total number of recordings. However, note that, since the \ac{qtdb} holds a wide variety of beat abnormalities, the percentage of very pathological or atypical beats is also relatively high. Hence, this number of abnormal beats detected is a positive result. 

The \ac{ecg} wave delineation results were compared to those of state-of-the-art algorithms, more specifically to multiscale parameter estimation \cite{Spicher2020}, low-complexity \ac{ecg} delineation \cite{Bote2018}, and wavelet-based methods \cite{Marco2011, Martinez2004}. Our approach achieved delineation results that were on a par with those of the other methods tested, as summarized in Table~\ref{tbl:del_results}. Further, the standard deviations of the differences between expert and algorithm annotations for the \ac{ecg} characteristic points of the P and T waves were lowest in our case, which indicates a very robust intra-recording delineation. This is confirmed by the results in Table~\ref{tbl:split_ds}, which shows that a very high percentage of recordings are in group I with small bias and standard deviation for the single \ac{ecg} characteristic points. Generally, we were able to achieve our goal: our new approach yields adequate \ac{ecg} wave delineation. 
\begin{table*}
\center
\caption{Comparing our work to state-of-the art methodologies for \ac{ecg} delination in terms of sensitivity, bias and standard deviation.}
\label{tbl:del_results}
\begin{tabular}{|c|c|c c c c c c c c|}
\hline
Method &  & P$_\text{on}$ & P$_\text{peak}$ & P$_\text{end}$ & QRS$_\text{on}$ & QRS$_\text{end}$ & T$_\text{on}$  & T$_\text{peak}$  & T$_\text{end}$ \\ \hline
       & \# beats & 3194 & 3194 & 3194 & 3626 & 3626 & 1412 & 3542 & 3542 \\ \hline \hline
\multirow{2}{*}{Our approach} & Se in \% & 98.5 & 98.5 & 98.5 & 100 & 100 & 100 & 100& 100 \\ 
 &  $\mu \pm \sigma$ in ms & 10.6 $\pm$ 12 & 7 $\pm$ 9 & 1.2 $\pm$ 11.7 & 6.7 $\pm$ 9.2 & 5.2 $\pm$ 10.2 & 2.1 $\pm$ 20.6 & 3.3 $\pm$ 11.9 & -5.6 $\pm$ 15.6 \\ \hline      
 
 \multirow{2}{*}{Spicher \cite{Spicher2020}} & Se in \% & 99.91 & 99.91 & 99.91 & 99.92 & 99.92 & 99.93 & 99.89& 99.89 \\ 
 &  $\mu \pm \sigma$ in ms & 0.5 $\pm$ 15.1 & 5.1 $\pm$ 10.9 & 0.5 $\pm$ 15.0 & 0.9 $\pm$ 8.5 & -0.4 $\pm$ 9.6 & 0.3 $\pm$ 23.7 & -4.5 $\pm$ 14.7 & 0.6 $\pm$ 20.3 \\ \hline   
 
 \multirow{2}{*}{Bote \cite{Bote2018}} & Se in \% & 98.22 & 99.34 & 99.87 & 100 & 99.97 & - & 99.89& 97.49 \\ 
 &  $\mu \pm \sigma$ in ms & 22.3 $\pm$ 14 & 13.5 $\pm$ 7.3 & -0.7 $\pm$ 9.5 & 7.0 $\pm$ 4.3 & -5 $\pm$ 9.9 & - & 8.4 $\pm$ 14.3 & -11.7 $\pm$ 15.0 \\ \hline  
 
  \multirow{2}{*}{Di Marco \cite{Marco2011}} & Se in \% & 98.15 & 98.15 & 98.15 & 100 & 100 & - & 99.72& 99.77 \\ 
 &  $\mu \pm \sigma$ in ms & -4.5 $\pm$ 13.4 & -4.7 $\pm$ 9.7 & -2.5 $\pm$ 13.0 & -5.1 $\pm$ 7.2 & 0.9 $\pm$ 8.7 & - & -0.3 $\pm$ 12.8 & 1.3 $\pm$ 18.6 \\ \hline  
  
  \multirow{2}{*}{Martinez \cite{Martinez2004}} & Se in \% & 98.87 & 98.87 & 98.75 & 99.97 & 99.97 & - & 99.77& 99.77 \\ 
 &  $\mu \pm \sigma$ in ms & 2 $\pm$ 14.8 & 4.8 $\pm$ 10.6 & 1.9 $\pm$ 12.8 & 4.6 $\pm$ 7.7 & 0.8 $\pm$ 8.7 & - & -0.2 $\pm$ 13.9 & -1.6 $\pm$ 18.1 \\ \hline  
   
\end{tabular}
\end{table*}

\begin{table*}[t]
\centering
\caption{Limits of $\mu_e$ and $\sigma_e$ for splitting the dataset given in the QT database into groups (left); corresponding performance comparison (values are given in \%) between our approach and our former approach \cite{Kovacs2017} (right).}
\vspace{1mm}
\label{tbl:split_ds}
\setlength\tabcolsep{2pt} 
\renewcommand{\arraystretch}{1.15}
\footnotesize
\begin{tabular}{|c|c|c|c|c|c|c|c|c|}
\hline 
&\multicolumn{2}{c|}{G I} & \multicolumn{2}{c|}{G II} & \multicolumn{2}{c|}{G III} & \multicolumn{2}{c|}{G IV}  \\
\hline 
 & $\mu_e$ & $\sigma_e$ & $\mu_e$ & $\sigma_e$ & $\mu_e$ & $\sigma_e$ & $\mu_e$ & $\sigma_e$ \\ 
\hline \hline
P$_\text{on}$ & $<$25 & $<$30 & $>$25 & $<$30 & $<$25 & $>$30 & $>$25 & $>$30 \\ 
\hline 
P$_\text{end}$ & $<$25 & $<$30 & $>$25 & $<$30 & $<$25 & $>$30 & $>$25 & $>$30 \\ 
\hline 
QRS$_\text{on}$ & $<$15 & $<$20 & $>$15 & $<$20 & $<$15 & $>$20 & $>$15 & $>$20 \\ 
\hline 
QRS$_\text{end}$ & $<$15 & $<$20 & $>$15 & $<$20 & $<$15 & $>$20 & $>$15 & $>$20 \\ 
\hline 
T$_\text{peak}$ & $<$40 & $<$50 & $>$40 & $<$50 & $<$40 & $>$50 & $>$40 & $>$50 \\ 
\hline 
T$_\text{peak}$ & $<$40 & $<$50 & $>$40 & $<$50 & $<$40 & $>$50 & $>$40 & $>$50 \\ 
\hline 
\end{tabular} 
%
%
\hfill
\label{tbl:cmp_opt}
\setlength\tabcolsep{1.8pt} 
\renewcommand{\arraystretch}{1.15}
\footnotesize
\begin{tabular}{|c||c|c|c|c||c|c|c|c||c|c|c|c|}
\hline 
\multirow{2}{*}{Method} & \multirow{2}{*}{G I} & \multirow{2}{*}{G II} & \multirow{2}{*}{G III} & \multirow{2}{*}{G IV}&  \multirow{2}{*}{G I}&  \multirow{2}{*}{G II} &  \multirow{2}{*}{G III} &  \multirow{2}{*}{G IV} &\multirow{2}{*}{G I}&  \multirow{2}{*}{G II} &  \multirow{2}{*}{G III} &  \multirow{2}{*}{G IV} \\
&  & & & & & & & & & & &  \\
\hline \hline
 & \multicolumn{4}{c||}{P$_\text{on}$} & \multicolumn{4}{c||}{P$_\text{end}$}& \multicolumn{4}{c|}{QRS$_\text{on}$} \\ 
\hline 
Our approach &93.68 & 5.26 & 1.05 & 0.00 & 93.68 & 3.16 & 1.05 & 2.11 & 88.57 & 6.67 & 2.86 & 1.90 \\
\hline 
Prv. approach \cite{Kovacs2017} &90.82 & 3.06 & 3.06 & 3.06 & 86.73 & 1.02 & 3.06 & 9.18 & 90.48 & 5.71 & 1.90 & 1.90 \\
\hline
 & \multicolumn{4}{c||}{QRS$_\text{end}$} & \multicolumn{4}{c||}{T$_\text{peak}$}& \multicolumn{4}{c|}{T$_\text{end}$} \\ 
\hline 
Our approach &87.62 & 6.67 & 2.86 & 2.86 & 94.17 & 2.91 & 1.94 & 0.97 & 95.15 & 2.91 & 0.97 & 0.97\\ 
\hline 
Prv. approach \cite{Kovacs2017} &86.67 & 6.67 & 2.86 & 3.81 & 89.32 & 5.83 & 0.97 & 3.88 & 95.15 & 2.91 & 0.97 & 0.97 \\
\hline 
\end{tabular}

\end{table*}

\section{Conclusion}
\label{sec:conclusion}

\begin{figure*}[!b]
\centering
\subfigure[Morphological changes related to translation of the T wave.]{
\includegraphics[width=5.5cm]{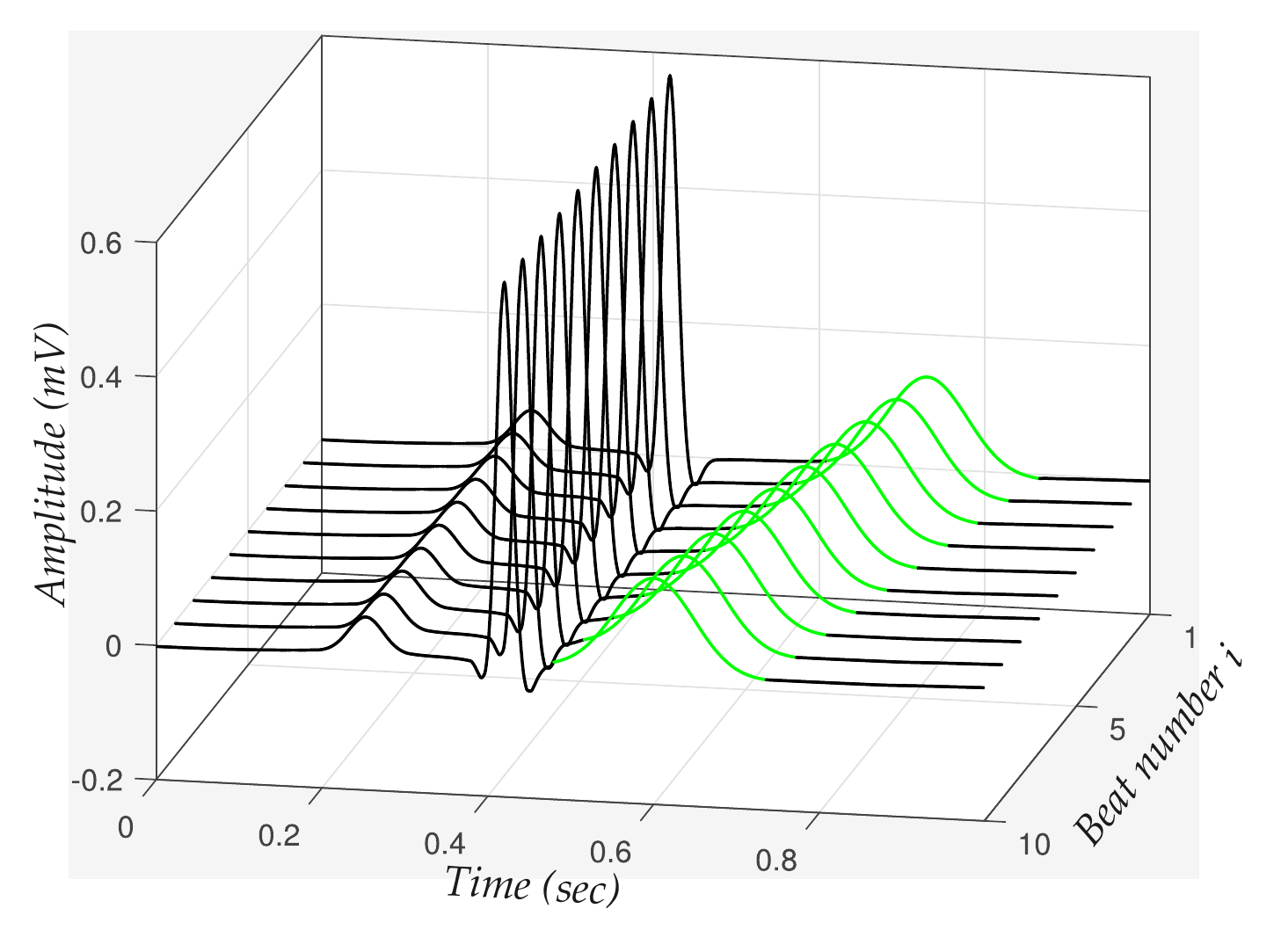}
\label{fig:trans_change}
}
\subfigure[Morphological changes related to dilation of the T wave.]{
\includegraphics[width=5.5cm]{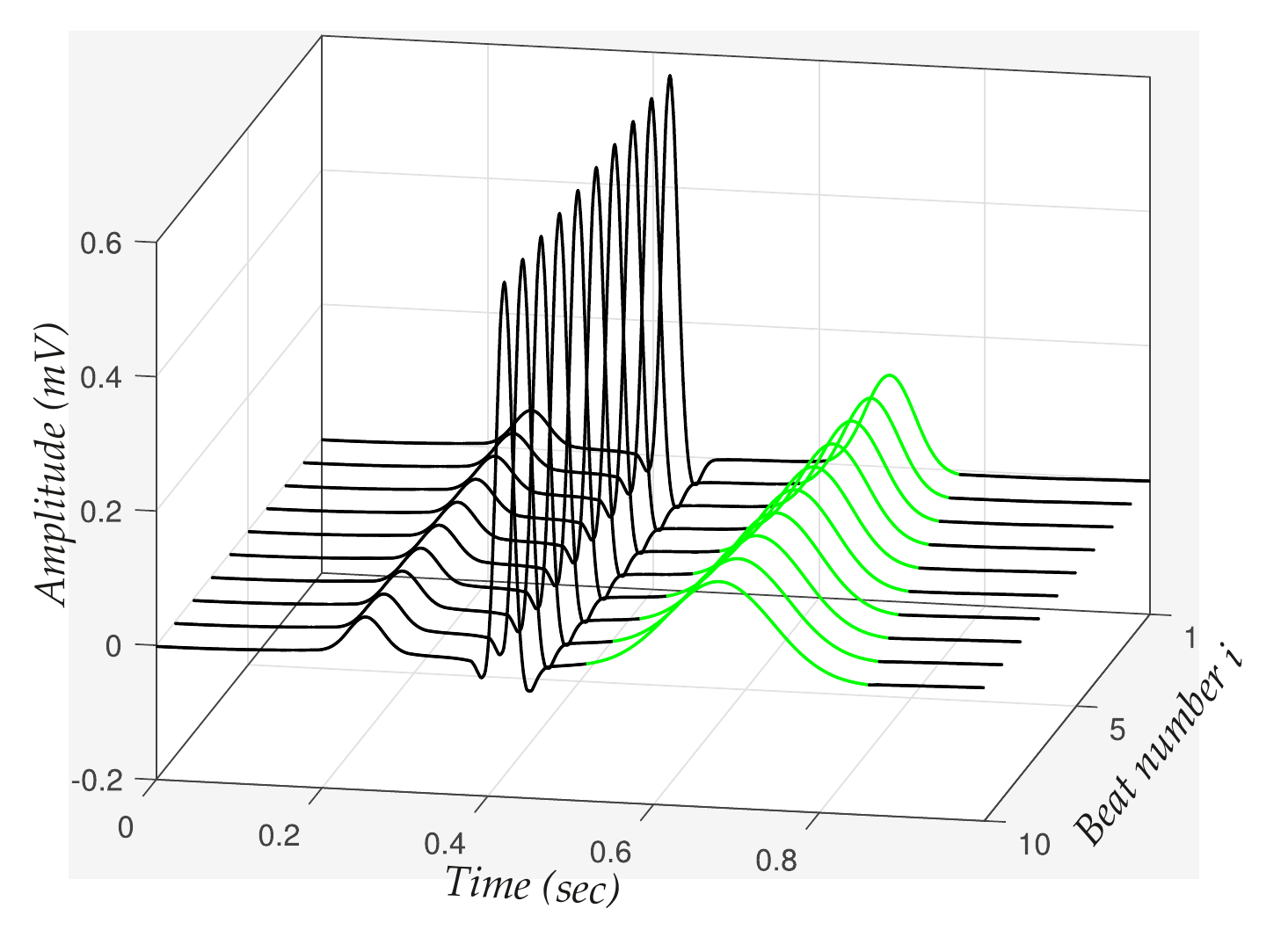}
\label{fig:dilat_change}
}
\subfigure[Actual morphological changes, i.e., positive T wave turning into a negative one.]{
\includegraphics[width=5.5cm]{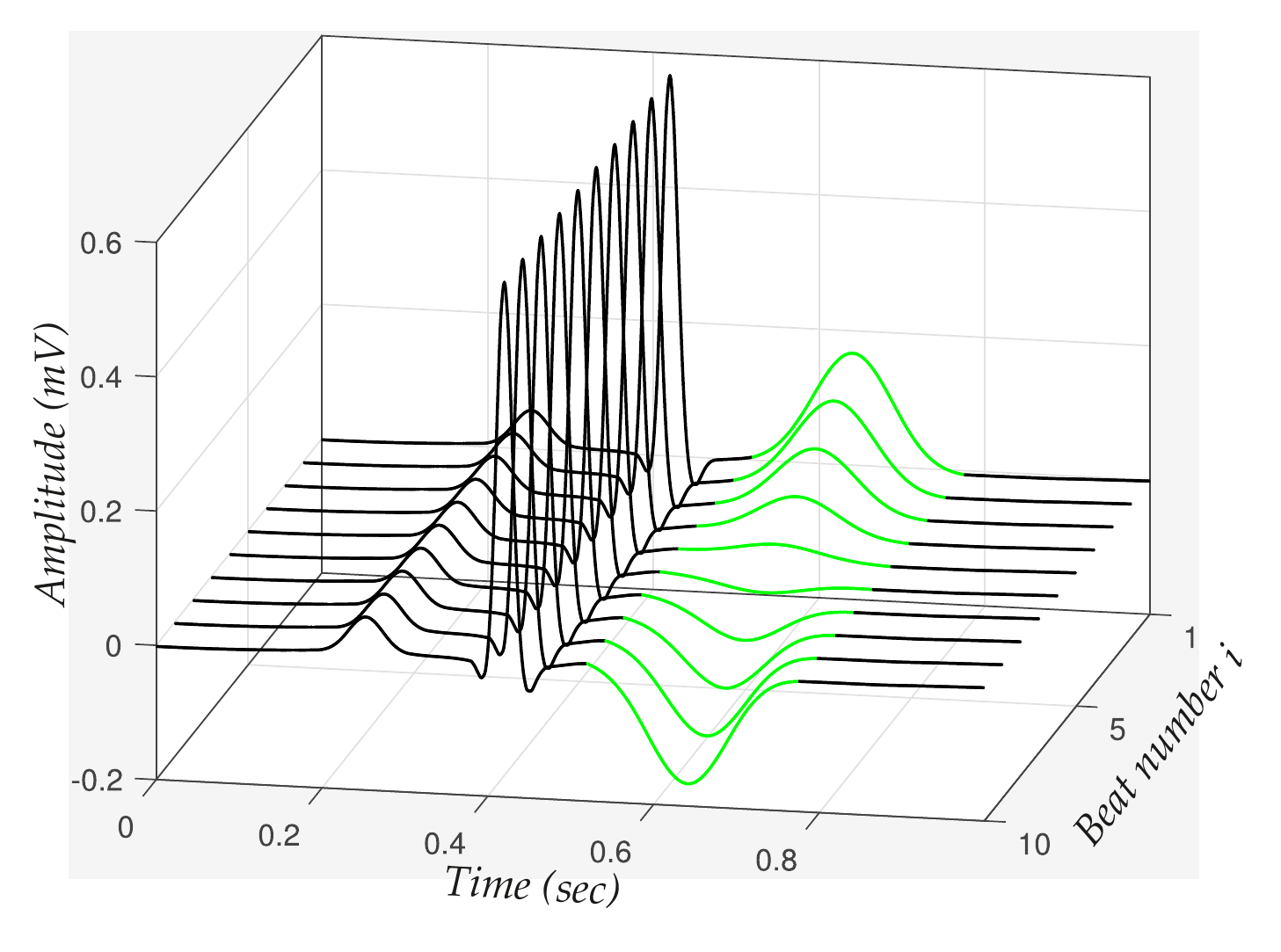}
\label{fig:morph_change}
} 
\caption{Three major effects which result in morphological changes of the \ac{ecg} beat or a single wave.}
\label{fig_intro}
\end{figure*}

We have illustrated that the combination of adaptive Hermite and sigmoidal functions with spline interpolation successfully copes with the challenges faced in \ac{ecg} signal processing. In particular, the ability of our method to properly perform \ac{ecg} \ac{blw} removal and \ac{ecg} delineation was demonstrated in Sections~\ref{sec:baseline} and \ref{sec:wave_segmentation}. These two tasks are fundamental to morphological information extraction -- the strength of our novel method. Morphological information extraction has several important applications in \ac{ecg} signal processing, for instance, distinguishing ischemic from non-ischemic ST changes \cite{Moody2003}, quantifying ventricular repolarization instability \cite{Laguna2016}, arrhythmia detection \cite{Gupta2020} or -- more generally -- evaluating the temporal evolution of the wave shapes. All these applications have in common that shape changes are related to different effects. Consequently, a distinction must be made between shape changes related to the translation of a wave (Fig.~\ref{fig:trans_change}), changes associated with the dilation of a wave (Fig.~\ref{fig:dilat_change}), and "actual" waveform changes, that is, a normally positive wave turning into a biphasic or even negative one (Fig.~\ref{fig:morph_change}). In real-world data, these changes usually occur in a super-positioned manner, which requires subsequent separation and identification. In fact, it is important to associate shape changes with their origins in the cardiovascular system to distinguish between those that are physiological and those that are pathological, otherwise specific biomedical signal couplings might be interpreted incorrectly and  important details may be overlooked. 

Our method is capable of differentiating between the distinct types of morphological variation and represents translation- and dilation-related nonlinear shape changes by $\tau$ and $\lambda$, respectively, while the coefficient vector $\mathbf{c}$ correlates with the "actual" linear wave shape changes.
Consequently, this work provides the mathematical concept underlying a novel method that combines and extends firmly established methods to encourage progress in research into morphology-based \ac{ecg} signal processing and analysis. The \textsc{MatLab} implementation of the proposed method is available at the website \cite{codes}.

\section*{Acknowledgment}

The authors would like to thank Axel Loewe and his team from the Karlsruhe Institute of Technology for providing simulated ECG data.

\ifCLASSOPTIONcaptionsoff
  \newpage
\fi



%
\newpage
\bibliographystyle{IEEEtr}
\bibliography{IEEEabrv,refs_paper}

\begin{thebibliography}{10}

\bibitem{sornmobook}
L.~Sörnmo and P.~Laguna, ``Chapter 7 - {ECG} signal processing,'' in {\em
  Bioelectrical Signal Processing in Cardiac and Neurological Applications},
  pp.~453 -- 566, Burlington: Academic Press, 2005.

\bibitem{ecg_book}
G.~D. Clifford {\em et~al.}, {\em Advanced Methods And Tools for {ECG} Data
  Analysis}.
\newblock Massachusetts, USA: Artech House, 2006.

\bibitem{Laguna2016}
P.~{Laguna} {\em et~al.}, ``Techniques for ventricular repolarization
  instability assessment from the {ECG},'' {\em Proceedings of the IEEE},
  vol.~104, no.~2, pp.~392--415, 2016.

\bibitem{Lenis2017}
G.~Lenis {\em et~al.}, ``{Comparison of Baseline Wander Removal Techniques
  considering the Preservation of {ST} Changes in the Ischemic {ECG} : A
  Simulation Study},'' {\em Comput. Math. Method. M.}, vol.~2017, 2017.

\bibitem{Romero2011}
D.~{Romero} {\em et~al.}, ``Depolarization changes during acute myocardial
  ischemia by evaluation of {QRS} slopes: Standard lead and vectorial
  approach,'' {\em {IEEE} Trans. Biomed. Eng.}, vol.~58, pp.~110--120, Jan
  2011.

\bibitem{Almeida2006}
R.~{Almeida} {\em et~al.}, ``{QT} variability and {HRV} interactions in {ECG}:
  quantification and reliability,'' {\em {IEEE} Trans. Biomed. Eng.}, vol.~53,
  no.~7, pp.~1317--1329, 2006.

\bibitem{Arini2008}
P.~D. Arini {\em et~al.}, ``{T}-wave width as an index for quantification of
  ventricular repolarization dispersion: Evaluation in an isolated rabbit heart
  model,'' {\em Biomed. Signal Proces.}, vol.~3, no.~1, pp.~67 -- 77, 2008.

\bibitem{Arini2014}
P.~D. Arini {\em et~al.}, ``Evaluation of ventricular repolarization dispersion
  during acute myocardial ischemia: spatial and temporal {ECG} indices,'' {\em
  Med. Biol. Eng. Comput.}, vol.~52, pp.~375--391, Apr 2014.

\bibitem{Kovacs2017}
P.~{Kovács} {\em et~al.}, ``{ECG} segmentation using adaptive {H}ermite
  functions,'' in {\em 51st Asilomar Conf. on Sign., Syst., and Comput.},
  pp.~1476--1480, Oct 2017.

\bibitem{hexp3}
P.~Laguna {\em et~al.}, ``Adaptive estimation of {QRS} complex by the {H}ermite
  model for classification and ectopic beat detection,'' {\em Med. and Biol.
  Eng. and Comp.}, vol.~3, pp.~58--68, 1996.

\bibitem{hexp5}
A.~Sandryhaila {\em et~al.}, ``Efficient compression of {QRS} complexes using
  {H}ermite expansion,'' {\em {IEEE} Trans. Signal Process.}, vol.~60, no.~2,
  pp.~947--955, 2012.

\bibitem{Kovacs2020}
P.~{Kovács} {\em et~al.}, ``Generalized rational variable projection with
  application in {ECG} compression,'' {\em IEEE Transactions on Signal
  Processing}, vol.~68, pp.~478--492, 2020.

\bibitem{hexp4}
M.~Lagerholm {\em et~al.}, ``Clustering {ECG} complexes using {H}ermite
  functions and self-organizing maps,'' {\em {IEEE} Trans. Biomed. Eng.},
  vol.~47, no.~7, pp.~838--717, 2000.

\bibitem{hexp1}
H.~Haraldsson {\em et~al.}, ``Detecting acute myocardial infarction in the
  12-lead {ECG} using {H}ermite expansions and neural networks,'' {\em Artif.
  Intell. in Med.}, vol.~32, pp.~127--136, 2004.

\bibitem{Ansari2017}
S.~{Ansari} {\em et~al.}, ``A review of automated methods for detection of
  myocardial ischemia and infarction using electrocardiogram and electronic
  health records,'' {\em IEEE Reviews in Biomedical Engineering}, vol.~10,
  pp.~264--298, 2017.

\bibitem{PhysioNet}
A.~L. Goldberger {\em et~al.}, ``{PhysioBank, PhysioToolkit, and PhysioNet}:
  Components of a new research resource for complex physiologic signals,'' {\em
  Circulation}, vol.~101, no.~23, pp.~215--220, 2000.

\bibitem{Laguna1997}
P.~Laguna {\em et~al.}, ``A database for evaluation of algorithms for
  measurement of {QT} and other waveform intervals in the {ECG},'' in {\em
  Comput. Cardiol. Conf.}, pp.~673--676, Sep 1997.

\bibitem{addison2005}
P.~S. {Addison}, ``Wavelet transforms and the {ECG}: a review,'' {\em
  Physiological measurement}, vol.~26, no.~5, p.~R155, 2005.

\bibitem{castells2007}
F.~{Castells} {\em et~al.}, ``Principal component analysis in {ECG} signal
  processing,'' {\em EURASIP Journal on Advances in Signal Processing},
  vol.~2007, pp.~1--21, 2007.

\bibitem{golub_pereyra1973}
G.~H. Golub and V.~Pereyra, ``{The differentiation of pseudo-inverses and
  nonlinear least squares problems whose variables separate},'' {\em SIAM J. on
  Numer. Anal.}, vol.~10, pp.~413--432, 1973.

\bibitem{varpro_matlab}
D.~P. O'Leary and B.~W. Rust, ``{Variable Projection for Nonlinear Least
  Squares Problems},'' {\em Comput. Optim. Appl.}, vol.~54, no.~3,
  pp.~579--593, 2013.

\bibitem{golub_pereyra2003}
G.~H. Golub and V.~Pereyra, ``{Separable nonlinear least squares: The variable
  projection method and its applications},'' {\em Inverse probl.}, vol.~19,
  no.~2, pp.~R1--R26, 2003.

\bibitem{hexp2}
L.~S\"ornmo {\em et~al.}, ``A method for evaluation of {QRS} shape features
  using a mathematical model for the {ECG},'' {\em {IEEE} Trans. Biomed. Eng.},
  vol.~28, pp.~713--717, 1981.

\bibitem{hexp6}
T.~D\'ozsa and P.~Kov\'acs, ``{ECG} signal compression using adaptive {H}ermite
  functions,'' {\em Advances in Intelligent Systems and Computing}, vol.~399,
  pp.~245--254, 2015.

\bibitem{szego}
G.~Szeg\H{o}, {\em Orthogonal polynomials}.
\newblock New York, USA: AMS Colloquium Publications, 3rd~ed., 1967.

\bibitem{monspline}
F.~N. Fritsch and R.~E. Carlson, ``Monotone piecewise cubic interpolation,''
  {\em SIAM J. on Numer. Anal.}, vol.~17, no.~2, pp.~238--246, 1980.

\bibitem{RIJNBEEK2014}
P.~R. Rijnbeek {\em et~al.}, ``Normal values of the electrocardiogram for ages
  16–90years,'' {\em J. Electrocardiol.}, vol.~47, no.~6, pp.~914 -- 921,
  2014.

\bibitem{Pan1985}
J.~{Pan} and W.~J. {Tompkins}, ``A real-time {QRS} detection algorithm,'' {\em
  IEEE Transactions on Biomedical Engineering}, vol.~BME-32, no.~3,
  pp.~230--236, 1985.

\bibitem{GUPTA2019}
V.~Gupta {\em et~al.}, ``R-peak detection using chaos analysis in standard and
  real time {ECG} databases,'' {\em IRBM}, vol.~40, no.~6, pp.~341 -- 354,
  2019.

\bibitem{Loewe2011}
A.~{Loewe} {\em et~al.}, ``Determination of optimal electrode positions of a
  wearable {ECG} monitoring system for detection of myocardial ischemia: A
  simulation study,'' in {\em 2011 Comput. Cardiol. Conf.}, pp.~741--744, Sep.
  2011.

\bibitem{Loewe2015}
A.~Loewe {\em et~al.}, ``{{ECG}-Based Detection of Early Myocardial Ischemia in
  a Computational Model : Impact of Additional Electrodes, Optimal Placement,
  and a New Feature for {ST} Deviation},'' {\em Biomed Res. Int.}, vol.~2015,
  p.~8 pages, 2015.

\bibitem{Romero2018}
F.~P. Romero {\em et~al.}, ``Baseline wander removal methods for {ECG} signals:
  A comparative study,'' 2018.

\bibitem{Alste1985}
J.~A. {Van Alste} and T.~S. {Schilder}, ``Removal of base-line wander and
  power-line interference from the {ECG} by an efficient {FIR} filter with a
  reduced number of taps,'' {\em {IEEE} Trans. Biomed. Eng.}, vol.~BME-32,
  pp.~1052--1060, Dec 1985.

\bibitem{Pottalla1990}
E.~W. Pottala {\em et~al.}, ``Suppression of baseline wander in the {ECG} using
  a bilinearly transformed, null-phase filter,'' {\em J Electrocardiol.},
  vol.~22, pp.~243 -- 247, 1990.
\newblock Proceedings of the Engineering Foundation Conference Computerized
  Interpretation of the Electrocardiogram XIV.

\bibitem{Meyer1977}
C.~Meyer and H.~Keiser, ``Electrocardiogram baseline noise estimation and
  removal using cubic splines and state-space computation techniques,'' {\em
  Comput. Biomed. Res.}, vol.~10, no.~5, pp.~459 -- 470, 1977.

\bibitem{Laguna1992}
P.~{Laguna} {\em et~al.}, ``Adaptive filtering of {ECG} baseline wander,'' in
  {\em 1992 14th Annual International Conference of the IEEE Eng. Med. Biol.
  Soc. Ann.}, vol.~2, pp.~508--509, Oct 1992.

\bibitem{Canan1998}
S.~{Canan} {\em et~al.}, ``A method for removing low varying frequency trend
  from {ECG} signal,'' in {\em 2nd Int. Conf. Biomed. Eng. Days}, pp.~144--146,
  May 1998.

\bibitem{Chouhan2007}
V.~S. {Chouhan} and S.~S. {Mehta}, ``Total removal of baseline drift from {ECG}
  signal,'' in {\em Int. Conf. on Comput.: Theor. Ap.}, pp.~512--515, March
  2007.

\bibitem{Lenis2014}
G.~{Lenis} {\em et~al.}, ``Post extrasystolic {T} wave change in subjectswith
  structural healthy ventricles - measurement and simulation,'' in {\em Comput.
  Cardiol. Conf. 2014}, pp.~1069--1072, Sep. 2014.

\bibitem{Wagner2009}
G.~S. Wagner {\em et~al.}, ``Aha/accf/hrs recommendations for the
  standardization and interpretation of the electrocardiogram: Part vi,'' {\em
  J. of the Am. Coll. Card.}, vol.~53, no.~11, pp.~1003 -- 1011, 2009.

\bibitem{Bennet2014}
M.~T. Bennett {\em et~al.}, ``{Effect of beta-blockers on {QT} dynamics in the
  long {QT} syndrome: measuring the benefit},'' {\em EP Europace}, vol.~16,
  pp.~1847--1851, 05 2014.

\bibitem{Gonzalez2017}
F.~{González} {\em et~al.}, ``The physionet {QT} database: Study on the
  reliability of {P}-wave manual annotations under noisy recordings,'' in {\em
  Comput. Cardiol. Conf. (CinC)}, pp.~1--4, Sep. 2017.

\bibitem{Laguna1994}
P.~{Laguna} {\em et~al.}, ``Automatic detection of wave boundaries in multilead
  {ECG} signals: Validation with the cse database,'' {\em Computers and
  Biomedical Research}, vol.~27, pp.~45--60, 1994.

\bibitem{Martinez2004}
J.~P. Martinez {\em et~al.}, ``A wavelet-based {ECG} delineator: evaluation on
  standard databases,'' {\em {IEEE} Trans. Biomed. Eng.}, vol.~51,
  pp.~570--581, April 2004.

\bibitem{Jane1997}
R.~Jane {\em et~al.}, ``Evaluation of an automatic threshold based detector of
  waveform limits in {H}olter {ECG} with the {QT} database,'' in {\em 1997
  Comput. Cardiol. Conf.}, pp.~295--298, Sep 1997.

\bibitem{Spicher2020}
N.~{Spicher} and M.~{Kukuk}, ``Delineation of electrocardiograms using
  multiscale parameter estimation,'' {\em {IEEE} J. Biomed. Health Inform.},
  pp.~1--1, 2020.

\bibitem{Bote2018}
J.~M. {Bote} {\em et~al.}, ``A modular low-complexity {ECG} delineation
  algorithm for real-time embedded systems,'' {\em {IEEE} J. Biomed. Health
  Inform.}, vol.~22, no.~2, pp.~429--441, 2018.

\bibitem{Marco2011}
L.~Y. {Di Marco} and L.~A. {Chiari}, ``A wavelet-based {ECG} delineation
  algorithm for 32-bit integer online processing,'' {\em Biomed. Signal Proces.
  Online}, vol.~10, no.~23, 2011.

\bibitem{Moody2003}
G.~B. {Moody} and F.~{Jager}, ``Distinguishing ischemic from non-ischemic {ST}
  changes: the physionet/computers in cardiology challenge 2003,'' in {\em
  Comput. Cardiol. Conf.}, pp.~235--237, 2003.

\bibitem{Gupta2020}
V.~{Gupta} {\em et~al.}, ``Chaos theory: An emerging tool for arrhythmia
  detection,'' {\em Sensing and Imaging}, vol.~21, no.~10, 2020.

\bibitem{codes}
C.~B\"ock and P.~Kov\'acs, ``{ECG} beat representation and delineation by means
  of variable projection,'' 2020, [{O}nline].
\newblock Available: http://codeocean.com.

\end{thebibliography}

%



\end{document}